\documentclass[a4paper,11pt]{article}

\usepackage[margin=2.5cm]{geometry}
\usepackage[T1]{fontenc}
\usepackage[utf8]{inputenc}

\usepackage[numbers,sort&compress]{natbib}

\usepackage{amsmath}        %
\usepackage{mathcommand}    %
\usepackage{mathtools}      %
\usepackage{xparse}         %

\makeatletter
\ifx\XeTeXrevision\undefined
  \usepackage{amsfonts}
\else
  \@ifpackageloaded{unicode-math}
    {
      \setmathfont{Latin Modern Math}[range=\mathcal]
    }
    {
      \usepackage[warnings-off={mathtools-overbracket}]{unicode-math}
      \setmathfont{Latin Modern Math}
    }
\fi
\makeatother

\makeatletter
\@ifpackageloaded{unicode-math}
  {
    \newcommand*{\fmtvector}[1]{{%
      \ifcat\relax\noexpand#1
        \symbfit{#1}%
      \else
        \symbfup{#1}%
      \fi
    }}
  }
  {
    \usepackage{bm}
    \newcommand*{\fmtvector}[1]{{%
      \ifcat\relax\noexpand#1
        \bm{#1}%
      \else
        \mathbf{#1}%
      \fi
    }}
  }
\makeatother

\providecommand{\vf}{\fmtvector{f}}

\providecommand{\vq}{\fmtvector{q}}
\providecommand{\vr}{\fmtvector{r}}
\providecommand{\vs}{\fmtvector{s}}
\providecommand{\vt}{\fmtvector{t}}
\providecommand{\vu}{\fmtvector{u}}
\providecommand{\vv}{\fmtvector{v}}
\providecommand{\vw}{\fmtvector{w}}
\providecommand{\vx}{\fmtvector{x}}
\providecommand{\vy}{\fmtvector{y}}

\providecommand{\vgamma}{\fmtvector{\gamma}}
\providecommand{\vdelta}{\fmtvector{\delta}}
\providecommand{\vepsilon}{\fmtvector{\epsilon}}
\providecommand{\vvarepsilon}{\fmtvector{\varepsilon}}
\providecommand{\vzeta}{\fmtvector{\zeta}}
\providecommand{\veta}{\fmtvector{\eta}}
\providecommand{\vtheta}{\fmtvector{\theta}}

\providecommand{\vmu}{\fmtvector{\mu}}

\providecommand{\vxi}{\fmtvector{\xi}}

\providecommand{\vsigma}{\fmtvector{\sigma}}

\providecommand{\vpsi}{\fmtvector{\psi}}

\providecommand{\vnull}{\fmtvector{0}}

\makeatletter
\@ifpackageloaded{unicode-math}
  {
    \newcommand*{\fmtmatrix}[1]{{\symbfup{#1}}}
  }
  {
    \usepackage{bm}
    \newcommand*{\fmtmatrix}[1]{{%
      \ifcat\relax\noexpand#1
        \bm{#1}%
      \else
        \mathbf{#1}%
      \fi
    }}
  }
\makeatother

\providecommand{\mC}{\fmtmatrix{C}}

\providecommand{\mH}{\fmtmatrix{H}}
\providecommand{\mI}{\fmtmatrix{I}}
\providecommand{\mJ}{\fmtmatrix{J}}
\providecommand{\mK}{\fmtmatrix{K}}

\providecommand{\mM}{\fmtmatrix{M}}

\providecommand{\mGamma}{\fmtmatrix{\Gamma}}

\providecommand{\mSigma}{\fmtmatrix{\Sigma}}

\providecommand{\mnull}{\fmtmatrix{0}}

\providecommand{\fmtset}[1]{{\mathbb{#1}}}

\providecommand{\sR}{\fmtset{R}}

\providecommand{\R}{\sR}   %

\newcommand*{\super}[2]{{#1^{#2}}}

\newcommand*{\transposesymbol}{\mathsf{T}}

\newcommand*{\transpose}[1]{\super{#1}{\transposesymbol}}

\newcommand*{\inverse}[1]{\super{#1}{-1}}

\providecommand{\Tinv}[1]{\super{#1}{-\transposesymbol}}

\providecommand{\invT}[1]{\Tinv}
\providecommand{\pinvT}[1]{\Tpseudoinv}

\DeclarePairedDelimiter{\parens}{\lparen}{\rparen}
\DeclarePairedDelimiter{\bracks}{\lbrack}{\rbrack}
\DeclarePairedDelimiter{\braces}{\lbrace}{\rbrace}
\DeclarePairedDelimiter{\angs}{\langle}{\rangle}
\DeclarePairedDelimiter{\singlepipes}{\lvert}{\rvert}
\DeclarePairedDelimiter{\doublepipes}{\lVert}{\rVert}

\newcommand*{\dif}[2][1]{\mathrm{d}\ifstrequal{#1}{1}{}{^{#1}} #2}
\newcommand*{\pdif}[2][1]{\partial\ifstrequal{#1}{1}{}{^{#1}} #2}
\newcommand*{\diff}[2][1]{\mathrm{d} #2\ifstrequal{#1}{1}{}{^{#1}}}
\newcommand*{\pdiff}[2][1]{\partial #2\ifstrequal{#1}{1}{}{^{#1}}}

\newcommand*{\integrateover}[3][]{\ifstrempty{#1}{\int}{\int_{#1}} {#2} \, \diff{#3}}

\NewDocumentCommand{\uniform}{s o m m}{
  \IfBooleanTF{#1}{
    \IfNoValueTF{#2}{
      \mathcal{U}\parens*{#3, #4}
    }{
      \mathcal{U}\parens*{#2 | #3, #4}
    }
  }{
    \IfNoValueTF{#2}{
      \mathcal{U}\parens{#3, #4}
    }{
      \mathcal{U}\parens{#2 | #3, #4}
    }
  }
}

\NewDocumentCommand{\normal}{s o m m}{
  \IfBooleanTF{#1}{
    \IfNoValueTF{#2}{
      \mathcal{N}\parens*{#3, #4}
    }{
      \mathcal{N}\parens*{#2 | #3, #4}
    }
  }{
    \IfNoValueTF{#2}{
      \mathcal{N}\parens{#3, #4}
    }{
      \mathcal{N}\parens{#2 | #3, #4}
    }
  }
}

\NewDocumentCommand{\dirac}{s m}{
  \IfBooleanTF{#1}
    {\delta\parens*{#2}}
    {\delta\parens{#2}}
}

\NewDocumentCommand{\GP}{s m m}{
  \IfBooleanTF{#1}
    {\mathcal{GP}\parens*{#2, #3}}
    {\mathcal{GP}\parens{#2, #3}}
}

\NewDocumentCommand{\GF}{s m m}{
  \IfBooleanTF{#1}
    {\mathcal{GF}\parens*{#2, #3}}
    {\mathcal{GF}\parens{#2, #3}}
}

\DeclareMathOperator{\variance}{Var}
\DeclareMathOperator{\covariance}{Cov}

\NewDocumentCommand{\probabilityof}{s o m}{
  \IfBooleanTF{#1}
    {p\parens*{#3\IfValueT{#2}{|#2}}}
    {p\parens{#3\IfValueT{#2}{|#2}}}
}

\NewDocumentCommand{\proposalof}{s m}{
  \IfBooleanTF{#1}
    {q\parens*{#2}}
    {q\parens{#2}}
}

\NewDocumentCommand{\expectationof}{s m}{
  \IfBooleanTF{#1}
    {\mathbb{E}\bracks*{#2}}
    {\mathbb{E}\bracks{#2}}
}

\NewDocumentCommand{\varianceof}{s m}{
  \IfBooleanTF{#1}
    {\variance\parens*{#2}}
    {\variance\parens{#2}}
}

\NewDocumentCommand{\covarianceof}{s m m}{
  \IfBooleanTF{#1}
    {\covariance\parens*{#2, #3}}
    {\covariance\parens{#2, #3}}
}

\DeclareMathOperator{\wasserstein}{W}

\NewDocumentCommand{\wassersteinof}{s o m m}{
  \IfBooleanTF{#1}
    {\wasserstein_{\IfValueTF{#2}{#2}{2}}\parens*{#3, #4}}
    {\wasserstein_{\IfValueTF{#2}{#2}{2}}\parens{#3, #4}}
}

\DeclareMathOperator*{\diag}{diag}
\DeclareMathOperator*{\argmax}{arg\,max}
\DeclareMathOperator*{\argmin}{arg\,min}
\DeclareMathOperator*{\linspan}{span}
\DeclareMathOperator*{\rank}{rank}
\DeclareMathOperator*{\trace}{tr}
\DeclareMathOperator*{\range}{range}

\NewDocumentCommand{\diagof}{s m}{
  \IfBooleanTF{#1}
    {\diag\parens*{#2}}
    {\diag\parens{#2}}
}

\NewDocumentCommand{\argmaxof}{s o m}{
  \IfBooleanTF{#1}
    {\argmax\IfValueT{#2}{_{#2}}\parens*{#3}}
    {\argmax\IfValueT{#2}{_{#2}}\parens{#3}}
}

\NewDocumentCommand{\argminof}{s o m}{
  \IfBooleanTF{#1}
    {\argmin\IfValueT{#2}{_{#2}}\parens*{#3}}
    {\argmin\IfValueT{#2}{_{#2}}\parens{#3}}
}

\NewDocumentCommand{\linspanof}{s m}{
  \IfBooleanTF{#1}
    {\linspan\parens*{#2}}
    {\linspan\parens{#2}}
}

\NewDocumentCommand{\rankof}{s m}{
  \IfBooleanTF{#1}
    {\rank\parens*{#2}}
    {\rank\parens{#2}}
}

\NewDocumentCommand{\traceof}{s m}{
  \IfBooleanTF{#1}
    {\trace\parens*{#2}}
    {\trace\parens{#2}}
}

\NewDocumentCommand{\rangeof}{s m}{
  \IfBooleanTF{#1}
    {\range\parens*{#2}}
    {\range\parens{#2}}
}

\NewDocumentCommand{\norm}{s o m}{
  \IfBooleanTF{#1}
    {\doublepipes*{#3}\IfValueT{#2}{_{#2}}}
    {\doublepipes{#3}\IfValueT{#2}{_{#2}}}
}

\NewDocumentCommand{\seminorm}{s o m}{
  \IfBooleanTF{#1}
    {\singlepipes*{#3}\IfValueT{#2}{_{#2}}}
    {\singlepipes{#3}\IfValueT{#2}{_{#2}}}
}

\NewDocumentCommand{\absolute}{s m}{
  \IfBooleanTF{#1}
    {\singlepipes*{#2}}
    {\singlepipes{#2}}
}

\NewDocumentCommand{\innerproduct}{s o m m}{
  \IfBooleanTF{#1}
    {\angs*{#3, #4}\IfValueT{#2}{_{#2}}}
    {\angs{#3, #4}\IfValueT{#2}{_{#2}}}
}

\NewDocumentCommand{\collection}{s m m m}{
  \IfBooleanTF{#1}
    {\braces*{#2, #3, \dots, #4}}
    {\braces{#2, #3, \dots, #4}}
}

\newcommand{\strain}{\varepsilon}          %
\newcommand{\strains}{\vvarepsilon}    %
\newcommand{\stress}{\sigma}            %
\newcommand{\stresses}{\vsigma}       %

\newcommand{\lame}{\lambda}          %
\newcommand{\shearmod}{\mu}          %
\newcommand{\poisson}{\nu}           %

\newcommand{\modulus}{E}                          %
\newcommand{\Einf}{\modulus_{\infty}}             %
\newcommand{\Eagg}{\modulus_{\mathrm{agg}}}       %
\newcommand{\modratio}{\rho}                       %
\newcommand{\lengthscale}{\ell}                    %

\newcommand{\axial}{z}                  %
\newcommand{\zitz}{\axial_{\mathrm{i}}} %
\newcommand{\force}{F}                   %

\newcommand{\params}{\vtheta}            %
\newcommand{\latent}{\vxi}               %
\newcommand{\latentmap}{\hat{\latent}}    %

\newcommand{\unodal}{\hat{\vu}}

\newcommand{\nspec}{G}                          %
\newcommand{\fwd}{\veta}                         %
\newcommand{\discrepancy}{\vdelta}              %
\newcommand{\noisevec}{\vepsilon}               %
\newcommand{\disckernel}{k_{\delta}}            %
\newcommand{\discvar}{\sigma_{\delta}^{2}}      %
\newcommand{\noisevar}{\sigma_{\epsilon}^{2}}   %
\newcommand{\corrlength}{\lambda_{\delta}}      %
\newcommand{\corrmat}{\mC}                       %
\newcommand{\obscov}{\mSigma}                    %
\newcommand{\hyperparams}{\vpsi}                 %

\newcommand{\refreshrate}{\lambda_{\mathrm{ref}}}  %

\newcommand{\descl}{\vzeta}                   %
\newcommand{\reparam}{\mathcal{T}}             %
\newcommand{\newdesc}{\tilde{\descl}}         %
\newcommand{\popmean}{\vmu}                    %
\newcommand{\popcov}{\mGamma}                  %
\newcommand{\popsd}{\gamma}                    %
\newcommand{\popsds}{\vgamma}                  %
\newcommand{\poppar}{(\popmean, \popcov)}      %
\newcommand{\alldata}{\vy_{1:\nspec}}          %

\newcommand{\rvedomain}{\Omega_{\mathrm{RVE}}}  %
\newcommand{\rvesize}{L}                         %
\newcommand{\naggr}{N}                            %
\newcommand{\macrostrains}{\strains^{M}}          %
\newcommand{\macrostressaxial}{\stress^{M}_{xx}}         %
\newcommand{\macrostresses}{\stresses^{M}}         %
\newcommand{\maxvonmises}{\stress^{\max}_{\mathrm{VM}}}         %
\newcommand{\maxprincipal}{\strain^{\max}_{I}}         %
\newcommand{\distfield}{d}                        %
\newcommand{\volavg}[1]{\langle #1 \rangle}      %
\newcommand{\qoi}{\vq}                            %
\newcommand{\rvemap}{\fwd_{\mathrm{RVE}}}        %
\newcommand{\blendweight}{w}                      %
\newcommand{\blendtemp}{\tau}                     %

\usepackage{xspace}

\makeatletter
\newcommand{\etal}{\@ifstar\etalstar\etalnostar}
\newcommand{\etc}{\@ifstar\etcstar\etcnostar}
\newcommand{\iid}{\@ifstar\iidstar\iidnostar}
\makeatother

\newcommand{\etalstar}{et~al.}
\newcommand{\etalnostar}{et~al.\@\xspace}
\newcommand{\ie}{i.e.\@\xspace}

\newcommand{\etcstar}{etc.}
\newcommand{\etcnostar}{etc.\@\xspace}
\newcommand{\iidstar}{i.i.d.}
\newcommand{\iidnostar}{i.i.d.\@\xspace}

\newcommand{\BPS}{Bouncy Particle sampler}
\newcommand{\PDMP}{PDMP}
\newcommand{\PDMPs}{PDMPs}

\newcommand{\PDE}{PDE}

\newcommand{\MAP}{MAP}
\newcommand{\GPabbr}{GP}

\newcommand{\ITZ}{ITZ}
\newcommand{\FEM}{FEM}
\newcommand{\RVE}{RVE}
\newcommand{\HCP}{hardened cement paste}
\newcommand{\XCT}{XCT}

\newcommand{\MC}{Monte Carlo}

\usepackage{graphicx}
\usepackage{amssymb}
\usepackage{booktabs}
\usepackage{multirow}
\usepackage{xcolor}
\usepackage{hyperref}

\usepackage{amsthm}%
\usepackage{textcomp}%
\usepackage{algorithm}%
\usepackage{algorithmicx}%
\usepackage{algpseudocode}%
\usepackage{listings}%

\usepackage{nicefrac}
\usepackage{microtype}
\usepackage{url}
\usepackage{xfrac}
\usepackage{import}
\usepackage[group-separator={,}]{siunitx}
\usepackage{subcaption}
\usepackage[colorinlistoftodos]{todonotes}
\presetkeys{todonotes}{inline}{}
\usepackage{textcase}
\usepackage{dsfont}
\usepackage{placeins}

\usepackage[capitalise]{cleveref}

\usepackage{tikz}
\usetikzlibrary{arrows.meta, fit, positioning}

\title{Probabilistic Upscaling of the Interfacial Transition Zone Behaviour in Concrete}

\author{%
  Leon Riccius\thanks{Corresponding author: \texttt{l.f.riccius@tudelft.nl}}
  \and Iuri B.C.M. Rocha
  \and Erik Schlangen
  \and Frans P. {van der Meer}\\[1ex]
  Faculty of Civil Engineering and Geosciences, Delft University of Technology,\\
  Stevinweg 1, 2628 CN Delft, The Netherlands
}
\date{}

\begin{document}

\maketitle

\begin{abstract}
The stiffness and cracking behaviour of concrete are influenced by the interfacial transition zone (\ITZ), the thin shell of paste that surrounds each aggregate.
Yet the \ITZ's mechanical properties are poorly characterized and are usually assigned to mesoscale models through ad hoc reduction factors.
Because the \ITZ\ is only tens of micrometres thick, its stiffness must be inferred indirectly from micromechanical tests, and the resulting estimates vary widely between studies and are rarely reported with a quantified uncertainty.
We present a workflow that calibrates a homogenized \ITZ\ model against the surface displacements of several micro-cantilever bending specimens and returns its stiffness descriptors as posterior distributions rather than single values.
The calibration is posed as a Bayesian inverse problem with a model-discrepancy term, sampled with a surrogate-assisted Bouncy Particle sampler, the first application of this sampler to a real-world engineering problem.
We compare a pooled model, which fits one descriptor vector to all specimens, with a hierarchical model that treats the specimens as draws from a common population and is sampled with a local extension of the sampler.
The hierarchical model separates the estimation uncertainty from the aggregate-to-aggregate material variability, where only the former shrinks as more data are added.
Propagated through a representative volume element of several aggregates, its predictive yields broader distributions of the mesoscale stress and strain and a wider spread variety of maximum von Mises stress locations.
The calibration places the \ITZ\ thickness above the \SIrange{30}{80}{\micro\meter} commonly assumed in the literature.
\end{abstract}


\section{Introduction}
\label{sec:introduction}

Concrete is heterogeneous over a range of length scales, and which of them a model resolves determines what it needs as input.
At the microscale, the individual phases of the hardened cement paste are distinguishable: anhydrous cement grains, the inner and outer hydration products that surround them, and the capillary voids between them.
Resolving this intricate microstructure is prohibitively expensive beyond the scale of a few grains, so models of concrete instead operate at the mesoscale, where the aggregates are resolved individually but the paste around them is homogenized into effective properties; at the macroscale even the aggregates disappear and the material is treated as homogeneous throughout.
A mesoscale model therefore stands or falls with the effective properties it assigns to the paste.
However, the paste is not the same everywhere.

The \HCP\ in a thin shell around each aggregate, the interfacial transition zone (\ITZ), has a microstructure that differs markedly from the bulk: it is more porous and richer in hydration products, which makes it the weakest constituent of the material and gives it a governing influence on the stiffness, cracking and durability of concrete.
Reported thicknesses range from \SIrange{30}{80}{\micro\meter} for regular Portland cement paste and river gravel~\cite{Zhang2019}.
A mesoscale model that resolves the aggregates must also resolve some form of \ITZ\ around them, and requires its mechanical properties as input.
Yet these properties remain poorly characterized: they are commonly assigned through ad hoc reduction factors applied to the bulk paste~\cite{Lei2022, Nitka2020, Wang2015, Grondin2014}, even though the predicted effective stiffness and fracture response are sensitive to the choice~\cite{Ramesh1996, Sharma2020, Xue2023, Zhou2021}.
The \ITZ\ descriptors that these models consume thus warrant a calibrated value with a credible range rather than a single assumed number.
The \ITZ s of alternative binder and aggregate types, which are increasingly adopted for improved sustainability, are even less well understood, reinforcing the need for a systematic calibration of the \ITZ\ properties of any material system of interest.

Measuring these properties directly is not an option.
Because the \ITZ\ is only tens of micrometres thick, its mechanical properties cannot be identified by conventional bulk tests and must instead be estimated indirectly from tests where the \ITZ\ dominates the response~\cite{Chen2024, Scrivener2004}.
Micro-cantilever bending and nanoindentation of cement-paste beams milled against an aggregate probe the relevant length scale~\cite{Zhang2019, Gan2018, Gan2020, Mondal2009, Xiao2013}, and direct interface tests provide complementary bond strengths~\cite{Jebli2018}.
These measurements are indirect: a load--displacement curve is not a stiffness, but must be interpreted through a mechanical model of the specimen, and the resulting estimates are scattered.

Inferring elastic properties from displacement or full-field measurements is well understood~\cite{Vigliotti2018, Fatehiboroujeni2020, Touminet2025, Jafari2025, Krijnen2026}.
Its Bayesian formulation treats the unknown properties as random variables and combines a prior with the likelihood of the observed data into a posterior distribution, so that the inferred quantities come with a full account of their uncertainty rather than a single point estimate.
Reducing a fully resolved microstructure to a few descriptors introduces a systematic model error, and a standard Bayesian formulation that ignores it risks overconfident estimates.
The model-discrepancy framework of Kennedy and O'Hagan~\cite{Kennedy2001a} was developed to absorb such error and has since been applied in structural identification~\cite{Kosikova2023}.
For the \ITZ, however, these tools have not been brought together: existing studies identify its stiffness by fitting a deterministic model to either the effective moduli of the composite~\cite{Hashin2002} or the response of micromechanical specimens~\cite{Zhang2019}, which yields a single set of properties with no account of either the measurement noise or the model error.

\begin{figure}[tbp]
  \centering
  \fontfamily{lmss}\selectfont
  \begin{tikzpicture}[
      img/.style={inner sep=0pt, outer sep=0pt},
      layer/.style={img, anchor=south west, fill=white, draw=black!45, line width=0.2pt},
      flow/.style={-{Stealth[length=2.2mm]}, line width=0.9pt, black!70},
      flowlbl/.style={font=\fontsize{6}{7}\selectfont, text=black, align=center, inner sep=2pt},
      ttl/.style={font=\scriptsize, anchor=base, inner sep=1pt},
      mini/.style={img, anchor=north},
      minilbl/.style={font=\fontsize{5}{6}\selectfont, anchor=north, inner sep=1pt},
    ]
    \def\ga{figures-graphical_abstract}
    \def\sidew{2.1cm}
    \def\midy{0.77}
    \def\ttly{2.05}
    \foreach \f/\k in {10/9, 09/8, 08/7, 07/6, 06/5, 05/4, 04/3, 03/2, 02/1, 01/0}
      \node[layer] at ({0.08*\k}, {0.065*\k}) {\includegraphics[width=\sidew]{\ga-side_views-\f}};
    \node[ttl] at (1.41, \ttly) {digital specimens};
    \node[img, draw=black!45, line width=0.2pt] (hom) at (7.4, \midy)
      {\includegraphics[width=\sidew]{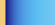}};
    \node[font=\footnotesize, anchor=north] at ([yshift=-1.5mm]hom.south)
      {\(\params = (\modratio, \lengthscale, \Einf)\)};
    \node[ttl] at (7.4, \ttly) {homogenized model};
    \node[img, anchor=west] (rve) at (10.45, \midy)
      {\includegraphics[width=2.2cm]{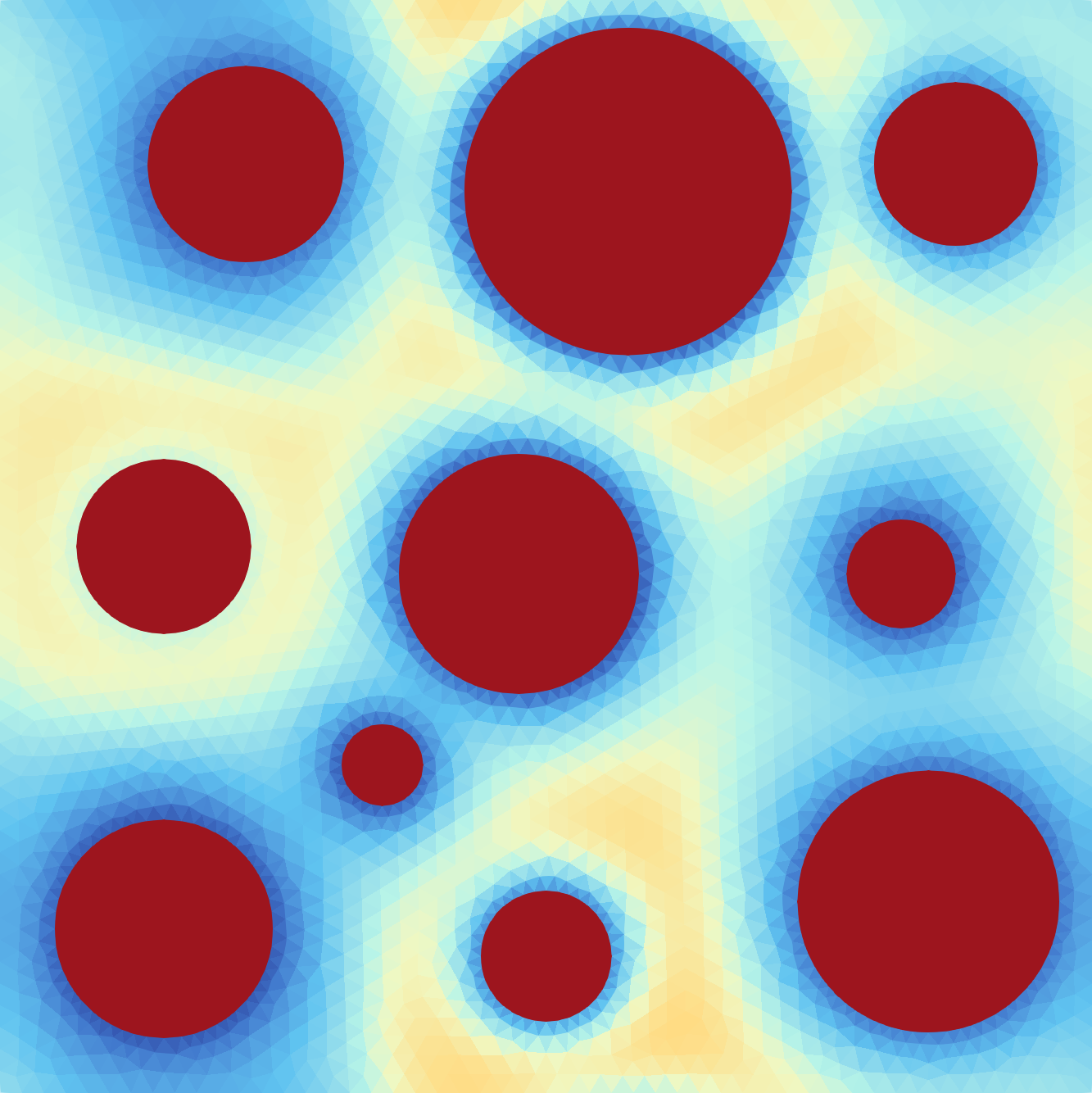}};
    \node[font=\footnotesize, anchor=north] at ([yshift=-0.5mm]rve.south)
      {\(\qoi = (\macrostressaxial, \maxvonmises, \maxprincipal)\)};
    \node[ttl] at (rve.north |- 0,\ttly) {\RVE};
    \draw[flow] (3.0, \midy) -- node[flowlbl, above] {calibration} (6.2, \midy);
    \foreach \name/\lbl [count=\i] in {point_estimate/\MAP, laplace_approximation/Laplace, sampling/sampling} {
      \node[mini] (m\i) at ({4.6 + 1.05*(\i-2)}, {\midy - 0.2})
        {\includegraphics[width=0.9cm]{\ga-\name}};
      \node[minilbl] at (m\i.south) {\lbl};
    }
    \draw[flow] (8.6, \midy) --
      node[flowlbl, above] {forward\\propagation}
      node[flowlbl, below] {linearization\\or \MC}
      (10.27, \midy);
  \end{tikzpicture}\par
  \vspace{4mm}
  \caption{
    Overview of the workflow.
    The homogenized \ITZ\ model is calibrated against the digital specimens, yielding the descriptors \(\params = (\modratio, \lengthscale, \Einf)\) as a point estimate (\MAP), a Laplace approximation or posterior samples.
    The calibrated descriptors are then propagated forward through a representative volume element (\RVE) of several aggregates, by linearization or by \MC, to the quantities of interest \(\qoi\): the axial homogenized stress \(\macrostressaxial\), the maximum von Mises stress \(\maxvonmises\) and the maximum principal strain \(\maxprincipal\).
  }
  \label{fig:graphical-abstract}
\end{figure}

This paper presents a workflow that calibrates a homogenized mesoscale model of the \ITZ\ against microscale displacement data while quantifying the uncertainty of the inferred descriptors (\cref{fig:graphical-abstract}).
The workflow applies whether the displacements are measured in a physical experiment or synthesized from microscale simulations on specimen geometries constructed from images, the former being the more demanding case.
We demonstrate it on synthesized data: a fully resolved finite-element model of the digital specimens of Zhang et al.~\cite{Zhang2019} supplies the displacements at a controlled, known noise level.
We describe the homogenized \ITZ\ by a small set of mesoscopic stiffness descriptors and infer them from the surface displacements of the micro-cantilever bending tests.

The simplest calibration is a point estimate, the maximum a posteriori (\MAP) descriptor vector, which we report for reference.
A single vector, however, gives no account of how well the data constrain it, nor of the model error discussed above.
We therefore sample the full posterior of the descriptors, with a likelihood that carries a Kennedy--O'Hagan model-discrepancy term to absorb the systematic error of the simplified \ITZ\ model.
Each evaluation of this likelihood and its gradient requires a finite-element solve, and a standard Markov chain Monte Carlo sampler needs many thousands of such evaluations.
To keep the inference affordable, we draw the posterior samples with the \BPS~\cite{Bouchard-Ct2018} through our surrogate-assisted piecewise-deterministic Markov process (\PDMP) framework for \PDE\ problems~\cite{Riccius2026b}, the first application of this sampler to a real-world engineering problem.

The micro-cantilever specimens from \citet{Zhang2019} differ appreciably in their mechanical response, a manifestation of the intrinsic spatial randomness of hardened cement paste~\cite{Tregger2007, Liu2024}.
A standard Bayesian model pools this data, fitting a single descriptor vector to all specimens at once.
Pooling is convenient, but is prone to conflate model error with the aggregate-to-aggregate variability of the material itself.
We therefore introduce a hierarchical model that treats the specimens as draws from a common population and infers that population.
It separates the two kinds of uncertainty that pooling conflates: what the finite data leave unresolved about the material and the aggregate-to-aggregate variability of the material itself.
Sampling the hierarchical posterior requires extending the surrogate-assisted \PDMP\ framework of~\cite{Riccius2026b}.
We exploit the hierarchical posteriors sparse structure to deploy a local version of the \BPS\ to keep the sampler affordable despite a much larger state space and expensive likelihood.

The calibrated posterior is a means to an end: what we ultimately want to know is how the inferred \ITZ\ shapes the mechanical response of concrete at the mesoscale, where many aggregates, each carrying its own \ITZ, sit close together and interact.
We therefore propagate the uncertainty in the descriptors through a representative volume element (\RVE) of several circular aggregates and obtain distributions of the resulting quantities of interest.
The hierarchical model matters most at this step: each aggregate in the volume element carries a fresh, unobserved \ITZ, so the propagation needs the predictive distribution of a new aggregate rather than one calibrated descriptor vector.
We carry both the pooled and the hierarchical model through the propagation, exposing how the modelling choice shapes the mesoscale fields and the upscaled properties.

\section{The interfacial transition zone specimen}
\label{sec:itz-specimen}

Our inference is driven by synthetic displacement measurements generated on digital specimens of the \ITZ\@.
We take these specimens from microscale imaging so that they carry a realistic microstructure with the porosity gradient and phase distribution of the real material near an aggregate.
This section describes their experimental origin, their reconstruction and segmentation from tomography, and the per-phase elastic properties that define the heterogeneous stiffness field of the forward model.

The digital specimens originate from the experimental and imaging campaign of \citet{Zhang2019}, whose micro-cantilever geometry we adopt as the reference microstructure for our forward model.
In their experiments, hardened cement paste (\mbox{w/c} ratio \num{0.3}, standard CEM~I \SI{42.5}{N} Portland cement) was cast onto one flat quartzite aggregate slice and cured for \num{28} days.
A micro-dicing saw then cut the paste layer in two perpendicular directions, leaving a row of \HCP\ blocks of square cross-section \qtyproduct{500 x 500}{\micro\meter} and length \SI{200}{\micro\meter}, each bonded at one end to the aggregate.

\begin{figure}[tbp]
  \centering
  {%
    \footnotesize
    \def\svgwidth{0.9\textwidth}
\begingroup%
  \makeatletter%
  \providecommand\color[2][]{%
    \errmessage{(Inkscape) Color is used for the text in Inkscape, but the package 'color.sty' is not loaded}%
    \renewcommand\color[2][]{}%
  }%
  \providecommand\transparent[1]{%
    \errmessage{(Inkscape) Transparency is used (non-zero) for the text in Inkscape, but the package 'transparent.sty' is not loaded}%
    \renewcommand\transparent[1]{}%
  }%
  \providecommand\rotatebox[2]{#2}%
  \newcommand*\fsize{\dimexpr\f@size pt\relax}%
  \newcommand*\lineheight[1]{\fontsize{\fsize}{#1\fsize}\selectfont}%
  \ifx\svgwidth\undefined%
    \setlength{\unitlength}{660bp}%
    \ifx\svgscale\undefined%
      \relax%
    \else%
      \setlength{\unitlength}{\unitlength * \real{\svgscale}}%
    \fi%
  \else%
    \setlength{\unitlength}{\svgwidth}%
  \fi%
  \global\let\svgwidth\undefined%
  \global\let\svgscale\undefined%
  \makeatother%
  \begin{picture}(1,0.54431818)%
    \lineheight{1}%
    \setlength\tabcolsep{0pt}%
    \put(0,0){\includegraphics[width=\unitlength,page=1]{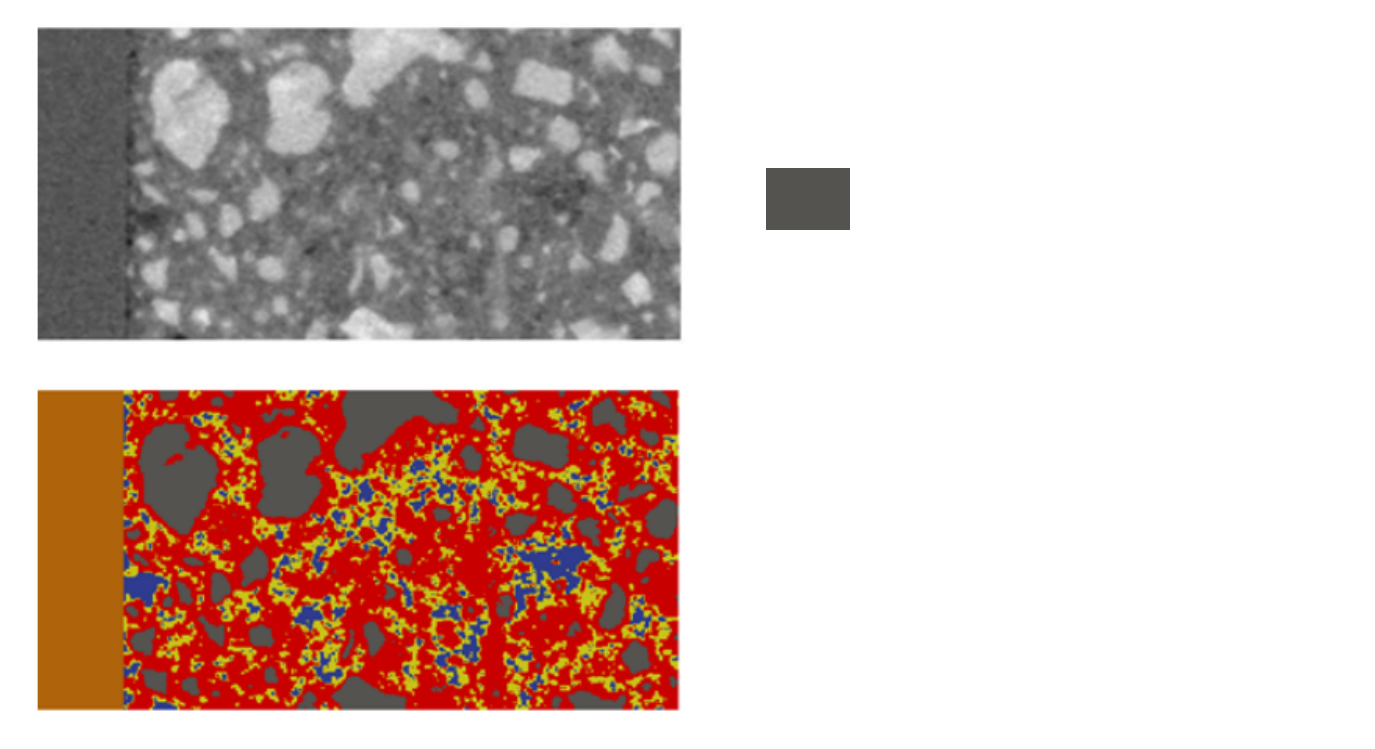}}%
    \put(0.63701963,0.39351631){\color[rgb]{0,0,0}\makebox(0,0)[lt]{\lineheight{0.1}\smash{\begin{tabular}[t]{l}Anhydrous cement particle (ACP)\end{tabular}}}}%
    \put(0,0){\includegraphics[width=\unitlength,page=2]{figures-specimen-x-ray-to-segmented.pdf}}%
    \put(0.63494443,0.32972967){\color[rgb]{0,0,0}\makebox(0,0)[lt]{\lineheight{0.1}\smash{\begin{tabular}[t]{l}Inner hydration product (IHP)\end{tabular}}}}%
    \put(0,0){\includegraphics[width=\unitlength,page=3]{figures-specimen-x-ray-to-segmented.pdf}}%
    \put(0.63584331,0.26594299){\color[rgb]{0,0,0}\makebox(0,0)[lt]{\lineheight{0.1}\smash{\begin{tabular}[t]{l}Outer hydration product (OHP)\end{tabular}}}}%
    \put(0,0){\includegraphics[width=\unitlength,page=4]{figures-specimen-x-ray-to-segmented.pdf}}%
    \put(0.63581002,0.20215635){\color[rgb]{0,0,0}\makebox(0,0)[lt]{\lineheight{0.1}\smash{\begin{tabular}[t]{l}Capillary Pore (CP)\end{tabular}}}}%
    \put(0,0){\includegraphics[width=\unitlength,page=5]{figures-specimen-x-ray-to-segmented.pdf}}%
    \put(0.63701963,0.13906882){\color[rgb]{0,0,0}\makebox(0,0)[lt]{\lineheight{0.1}\smash{\begin{tabular}[t]{l}Aggregate\end{tabular}}}}%
  \end{picture}%
\endgroup%

  }
  \caption{
    Reconstruction and segmentation of a paste--aggregate cross-section.
    Top: the grey-scale \XCT\ volume.
    Bottom: the corresponding five-phase segmentation: anhydrous cement particle, inner hydration product, outer hydration product, capillary pore, and aggregate.
    The aggregate occupies the left side; the heterogeneous hardened cement paste lies to its right.
    Adapted from \citet{Zhang2019}.
  }
  \label{fig:xct-segmentation}
\end{figure}

\citet{Zhang2019} reconstructed the blocks from X-ray computed tomography (\XCT) at resolution of \SI{2}{\micro\meter} voxels.
A \qtyproduct{300 x 300 x 1000}{\micro\meter} region of interest was then taken from the centre of the prism and segmented into five phases: the quartzite aggregate, and the four constituents of the paste -- capillary pore, outer hydration product, inner hydration product, and anhydrous cement particle.
Figure~\ref{fig:xct-segmentation} shows a cross-section of the reconstructed grey-scale volume alongside its segmentation, with the aggregate stub on the left and the heterogeneous paste, including the elevated near-interface porosity of the \ITZ, on the right.
Phase thresholds were calibrated on bulk paste more than \SI{150}{\micro\meter} from the aggregate, so that the porosity gradient of the \ITZ\ itself does not bias the segmentation.

We create 10 voxelized cantilever beams of \qtyproduct{100 x 100 x 220}{\micro\meter} on a \(50\times50\times110\) grid, comprising \SI{20}{\micro\meter} of aggregate and \SI{200}{\micro\meter} of paste.
The ten specimens are independent extractions from this segmented region of interest, and together they capture the variability of the paste microstructure across nominally identical beams.
Each voxel is assigned the phase it was segmented into, and each phase is given a uniform Young's modulus, listed in Table~\ref{tab:phase-moduli}; capillary pores are treated as void and carry no material.
A single Poisson's ratio of \num{0.18} is assumed for all solid phases.
These per-phase moduli define the heterogeneous stiffness field of the forward model.
We can then solve the governing equations of linear elasticity on the voxelized mesh, which yields the displacement field.
Finally, we can generate synthetic measurements based on this displacement field.

\begin{table}[t]
  \centering
  \caption{Phases of the digital \ITZ\ specimen and their assigned Young's moduli~\cite{Zhang2019}.
  Capillary pores are treated as void.}
  \label{tab:phase-moduli}
  \begin{tabular}{lc}
    \toprule
    Phase & Young's modulus (\si{\giga\pascal}) \\
    \midrule
    Aggregate (quartzite)     & 70 \\
    Anhydrous cement particle & 99 \\
    Inner hydration product   & 31 \\
    Outer hydration product   & 25 \\
    Capillary pore            & --- \\
    \bottomrule
  \end{tabular}
\end{table}

\section{The forward model}
\label{sec:forward-model}

The inference rests on two models of the same micro-cantilever bending test.
We generate the synthetic measurements with a fully resolved model that carries the heterogeneous microstructure of the digital specimen, and we calibrate against them a homogenized model in which the \ITZ\ is reduced to a smooth stiffness gradient described by a few parameters.
The resolved model fixes what the data are; the homogenized model is the object of the inference, and its parameters are the mesoscopic descriptors we set out to quantify.
We describe the resolved model first and then its homogenized counterpart, both sharing the same dimensions, loading, and observed quantities.

\subsection{Resolved heterogeneous model}
\label{sec:resolved-model}

Each digital specimen of \cref{sec:itz-specimen} is turned directly into a finite-element mesh, with one trilinear hexahedral element per solid voxel.
Capillary pores are treated as void and their voxels are removed from the mesh, so the mesh inherits the realistic, porous microstructure.
At this stage, the only approximation is the discretization of the geometry with its stepwise boundaries between phases.
Every remaining element is assigned the Young's modulus of the phase it was segmented into, listed in \cref{tab:phase-moduli}, giving a piecewise-constant modulus field \(\modulus(\vx)\) over the beam domain \(\Omega\).
A single Poisson's ratio \(\poisson = \num{0.18}\) is used for all solid phases.

We assume small strains and isotropic linear elasticity.
The displacement field \(\vu\) satisfies the equilibrium equation
\begin{equation}
  \nabla \cdot \stresses = \mnull \quad \text{in } \Omega,
  \label{eq:equilibrium}
\end{equation}
with the constitutive and kinematic relations
\begin{equation}
  \stresses = \lame \, \mathrm{tr}(\strains)\, \mI + 2\shearmod\, \strains,
  \qquad
  \strains = \tfrac{1}{2}\bigl(\nabla \vu + \transpose{(\nabla \vu)}\bigr),
  \label{eq:constitutive}
\end{equation}
where \(\vu\) are the displacements, \(\stresses\) is the Cauchy stress, \(\strains\) the small-strain tensor, \(\mI\) the identity, and the Lam\'e parameters \(\lame\) and \(\shearmod\) follow from the local modulus \(\modulus(\vx)\) and \(\poisson\).

\begin{figure}[tbp]
  \centering
  \includegraphics[width=.6\textwidth]{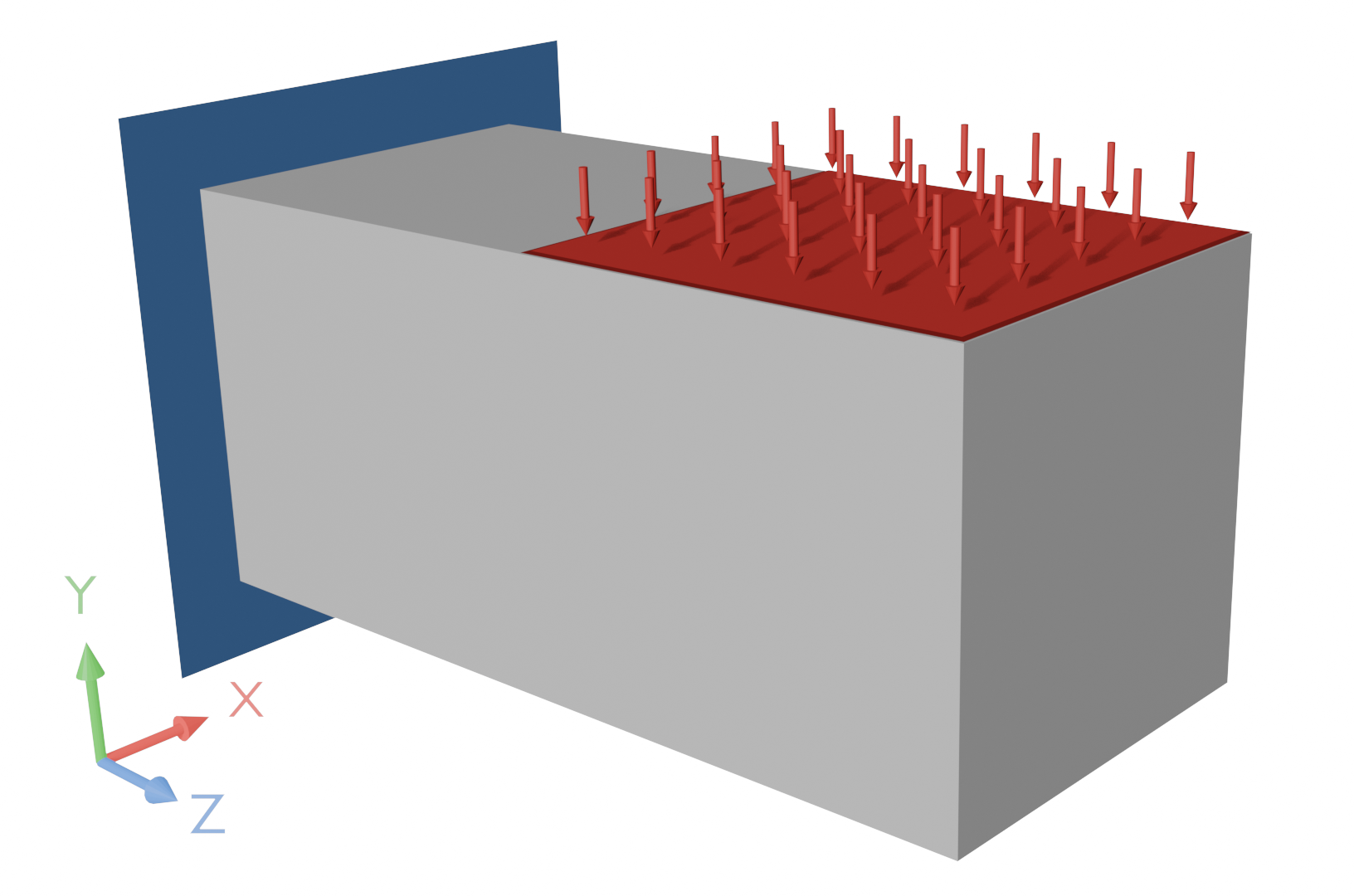}
  \caption{Cantilever beam outline with zero-displacements on the left side (blue base) and distributed load on the top lateral face (red).}
  \label{fig:beam-with-bc}
\end{figure}

The beam is loaded as a cantilever, analogous to the micro-cantilever tests of \citet{Zhang2019}, where a nano-indenter presses against one lateral face.
Let \(\axial\) denote the axial coordinate, measured from the aggregate base at \(\axial = 0\) to the free end at \(\axial = \SI{220}{\micro\meter}\).
The base is clamped, \(\vu = \vnull \), representing the bond to the bulk aggregate.
The indenter is represented by a uniform traction in the \(y\)-direction, transverse to the beam axis, applied over the free-end portion (\(\axial > \SI{110}{\micro\meter}\)) of one lateral face; its magnitude is set so that the resultant equals the prescribed force \(\force = \SI{60}{\milli\newton}\), which bends the beam about its clamped base.
All remaining faces are traction-free.
\Cref{fig:beam-with-bc} shows the beam outline with these boundary conditions.

We solve \crefrange{eq:equilibrium}{eq:constitutive} with the finite-element method.
Multiplying \cref{eq:equilibrium} by a test function \(\vw\) and integrating by parts over \(\Omega\) yields the weak form: find the displacement \(\vu\), with \(\vu = \vnull\) on the clamped base, such that
\begin{equation}
  \integrateover[\Omega]{\stresses(\vu) : \nabla \vw}{\Omega} = \integrateover[\Gamma_{\mathrm{N}}]{\vt \cdot \vw}{\Gamma}
  \label{eq:weak-form}
\end{equation}
holds for all test functions \(\vw\) that vanish on the clamped base, where \(\vt\) is the traction on the loaded surface \(\Gamma_{\mathrm{N}}\).
We discretize \(\vu\) and \(\vw\) with trilinear shape functions on the hexahedral elements, one per voxel, so that both fields are interpolated from their values at the element nodes.
Inserting this discretization into \cref{eq:weak-form} reduces the weak form to the linear system
\begin{equation}
  \mK \unodal = \vf,
  \label{eq:linear-system}
\end{equation}
where \(\unodal\) collects the nodal displacements, \(\mK\) is the stiffness matrix assembled from the element contributions, and \(\vf\) is the load vector produced by the traction.
This system of equations can then be solved for the nodal displacements \(\unodal\), which are interpolated to yield the continuous displacement field \(\vu\) over the beam.

\begin{figure}[tbp]
  \centering
  \subcaptionbox{Young's modulus \(\modulus(\vx)\)\label{fig:resolved-modulus}}{%
    \includegraphics[width=0.5\textwidth]{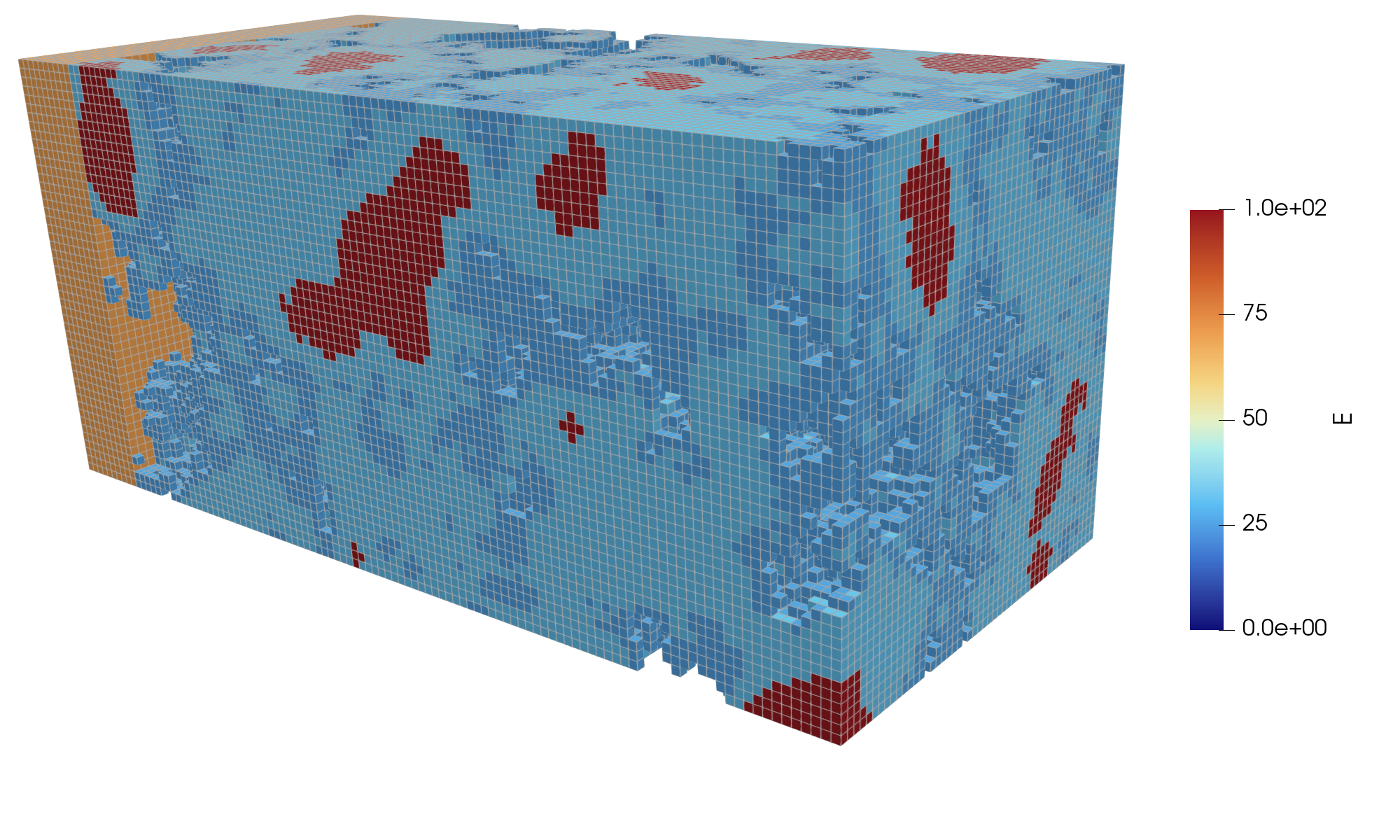}}%
  \subcaptionbox{Displacement field \(\vu\)\label{fig:resolved-displacement}}{%
    \includegraphics[width=0.5\textwidth]{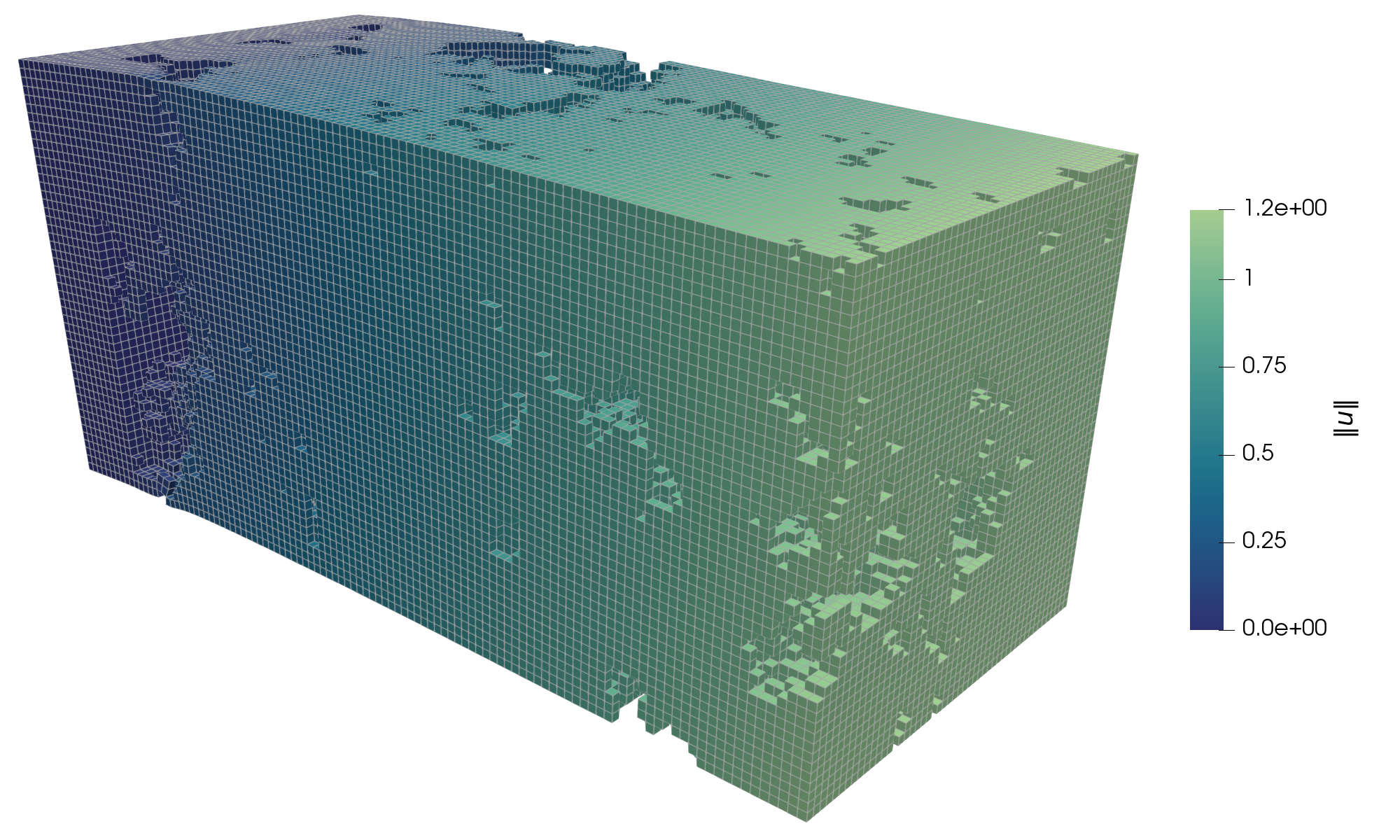}}%
  \caption{Resolved heterogeneous model of one digital specimen. a The piecewise-constant Young's modulus field, with each voxel carrying the modulus of its segmented phase and capillary pores removed as void. b The displacement field obtained by solving FEM model, amplified by factor \num{15}.}
  \label{fig:resolved-fields}
\end{figure}

We do not observe the full displacement field.
Instead, we record the displacements at a set of sensor points on the outer surface of the beam, mimicking the measurement of surface displacements.
These sensor displacements, corrupted by measurement noise, form the data vector \(\vy\) that drives the inference.
Each of the ten specimens of \cref{sec:itz-specimen} yields its own dataset, generated on an independent realization of the microstructure.
\Cref{fig:resolved-fields} shows the resolved modulus field of one specimen and the displacement field it produces under the cantilever load.

\subsection{Homogenized model}
\label{sec:homogenized-model}

We now replace the heterogeneous microstructure by a homogenized continuum in which the spatial variation of stiffness is reduced to a smooth gradient along the beam axis.
For the homogenized model, the geometry, the boundary value problem of \crefrange{eq:equilibrium}{eq:constitutive}, the loading, and the Poisson's ratio are kept identical; only the voids and other microstructural features are smoothed into a continuous modulus field .

In the homogenized model the Young's modulus is a smooth function of the axial coordinate alone, \(\modulus(\axial)\).
Over the aggregate stub (\(\axial < \zitz\), with \(\zitz = \SI{20}{\micro\meter}\)) it takes the constant aggregate value \(\Eagg = \SI{70}{\giga\pascal}\).
In the paste (\(\axial \ge \zitz\)) it recovers exponentially from a reduced near-interface value toward the bulk-paste modulus,
\begin{equation}
  \modulus(\axial) = \Einf \left[ 1 - (1 - \modratio)\, \exp\!\left(-\frac{\axial - \zitz}{\lengthscale}\right) \right],
  \qquad \axial \ge \zitz,
  \label{eq:recovery-field}
\end{equation}
where \(\Einf\) is the bulk-paste Young's modulus far from the interface, \(\modratio \in (0, 1]\) is the ratio of the interface modulus to the bulk value, so that \(\modulus(\zitz) = \modratio\, \Einf\), and \(\lengthscale\) is the length over which the modulus relaxes to the bulk.
This form encodes the porosity-driven stiffness gradient of the \ITZ: a soft, porous layer at the aggregate boundary stiffening to the bulk paste over a distance comparable to the \ITZ\ thickness.

\begin{figure}[tbp]
  \centering
  \subcaptionbox{Young's modulus \(\modulus(\axial)\)\label{fig:homogenized-modulus}}{%
    \includegraphics[width=0.5\textwidth]{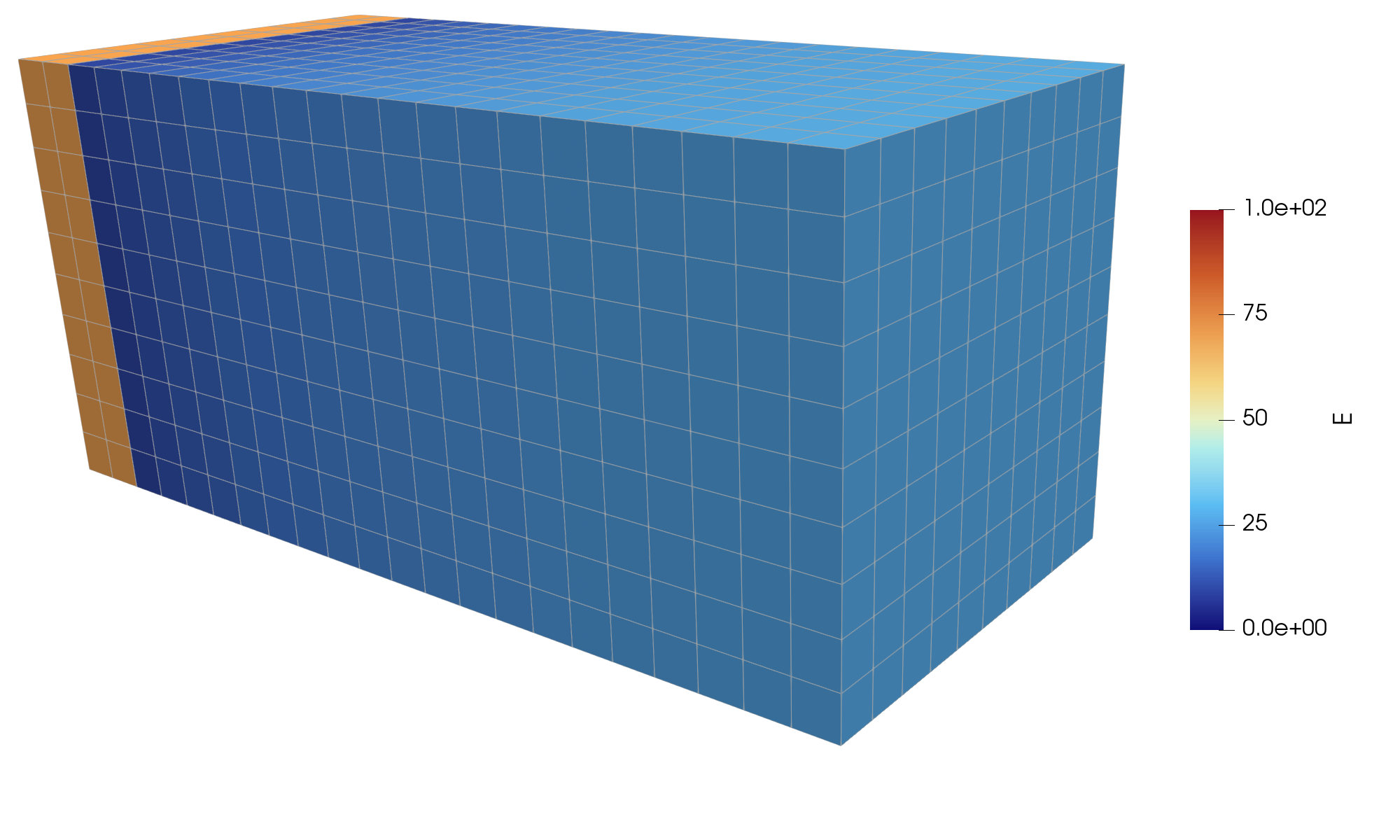}}%
  \subcaptionbox{Displacement field \(\vu\)\label{fig:homogenized-displacement}}{%
    \includegraphics[width=0.5\textwidth]{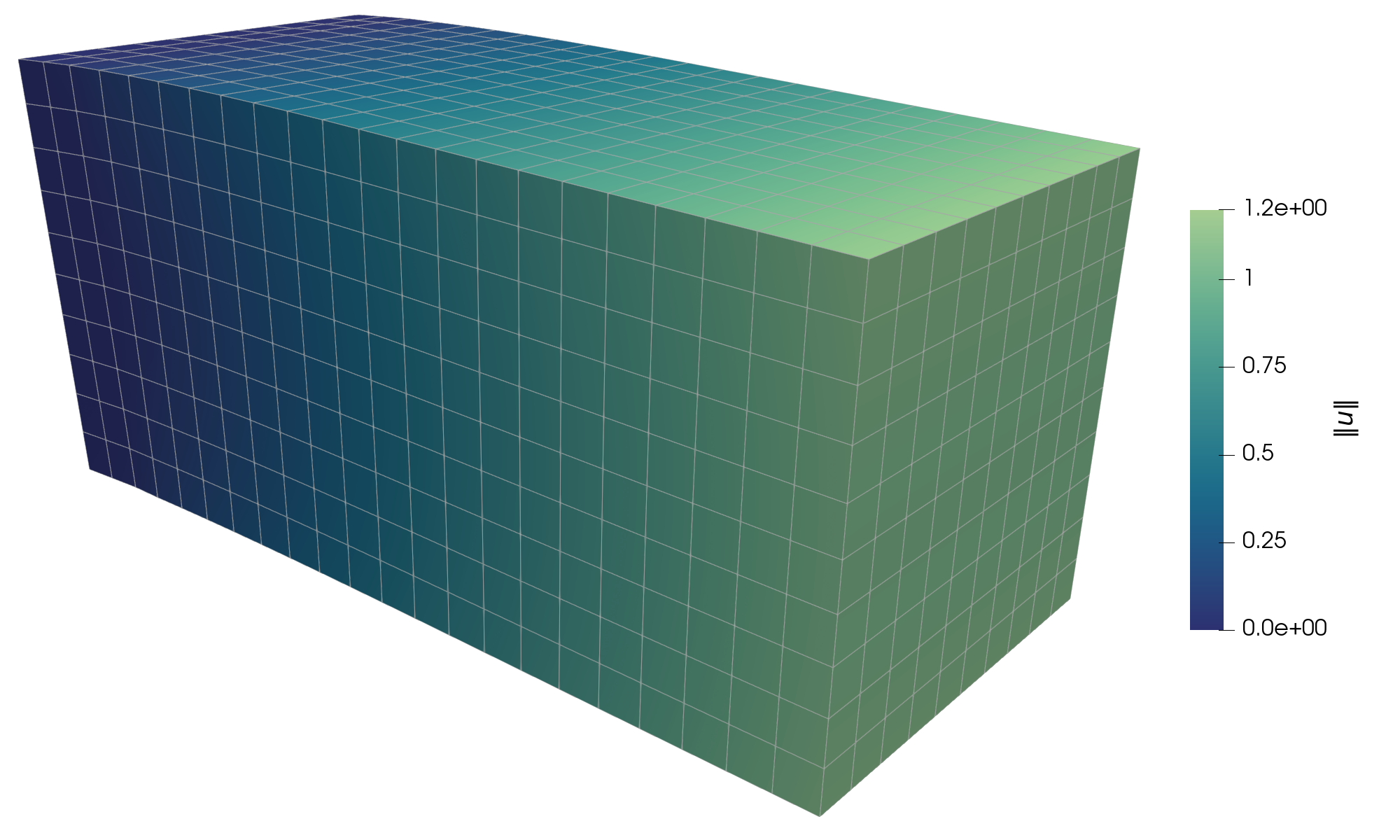}}%
  \caption{Homogenized model on the same geometry, loading, and boundary conditions as \cref{fig:resolved-fields}. a The Young's modulus reduced to a smooth gradient along the beam axis following \cref{eq:recovery-field}, constant over the aggregate stub and recovering exponentially toward the bulk-paste value in the paste. b The resulting displacement field on the coarse mesh, amplified by factor \num{15}.}
  \label{fig:homogenized-fields}
\end{figure}

The three parameters \(\params = (\modratio, \lengthscale, \Einf)\) are the unknowns of the inverse problem and the mesoscopic descriptors of the \ITZ\ we aim to quantify: \(\modratio\) measures how strongly the interface is weakened, \(\lengthscale\) how far the disturbance extends, and \(\Einf\) the bulk-paste stiffness.
Given \(\params\), the homogenized model defines the forward map \(\vy = \fwd(\params)\) that returns the displacements at the same sensor locations as the resolved model, so that model and data are compared directly in the likelihood.
Because it carries no microstructural features to resolve, the homogenized model is discretized on a much coarser mesh, with element size \SI{10}{\micro\meter}, which makes the repeated forward solves of the inference affordable.
\Cref{fig:homogenized-fields} shows the smooth axial modulus gradient and the displacement field of the homogenized counterpart to the specimen of \cref{fig:resolved-fields}.

We do not argue that the form of \cref{eq:recovery-field} is the only or best way to capture the \ITZ\ stiffness gradient, but it is a simple and interpretable choice that suffices to demonstrate the inference framework.
Depending on the specific application, other forms could be used, and the inference would proceed in the same way.
The systematic mismatch between this reduced description and the resolved data should naturally not be ignored and is therefore accounted for explicitly through a model-discrepancy term, introduced together with the likelihood below.

\subsection{Synthetic measurements}
\label{sec:synthetic-data}
The data are produced by the resolved model of \cref{sec:resolved-model}.
For each of the \(\nspec = 10\) digital specimens we solve the resolved boundary value problem on its segmented microstructure, under the cantilever loading of \cref{sec:resolved-model}, and read the resulting displacements at a set of sensor points on the outer surface.
The sensors are placed at random on the solid portion of the four lateral faces and the free-end face, drawn uniformly over the exposed surface and rejected where they fall on a pore, so that their density follows the solid fraction of each face.
Each sensor records the three Cartesian components of the surface displacement.
We corrupt every component with independent zero-mean Gaussian noise of standard deviation \(\sigma_{\epsilon} = \SI{0.01}{\micro\meter}\), amounting to about \SI{1}{\percent} of the maximum displacement.
Each of the \(\nspec\) specimens contributes its full set of \num{20} sensors, that is \num{60} observations per specimen and \num{600} observations in total.

\section{Likelihood and Bayesian inference}
\label{sec:inference}

The homogenized model of \cref{sec:homogenized-model} maps a descriptor vector \(\params\) to displacements at the sensor locations, and we identify \(\params\) from the measured displacements by Bayesian inference.
Bayes' theorem gives the posterior density of the unknowns, given the data, as proportional to the likelihood of the data times the prior density assigned to the unknowns before the data are seen.

Two effects separate the model's prediction from the measured data and so shape that likelihood.
The measurements carry observation noise, and the homogenized model is a deliberate simplification of the resolved microstructure, so that even at its best-fitting parameters its predictions differ systematically from the data.
We account for both with a likelihood that combines independent measurement noise with a structured model-discrepancy term following Kennedy and O'Hagan~\cite{Kennedy2001a}, defined in \cref{sec:likelihood}.

Beyond the likelihood itself, two choices remain, and the rest of the section works through each in turn.
The first is how the specimens enter it, and the second is how we characterize the resulting posterior.
Pooling them constrains a single descriptor vector \(\params\) to explain every specimen at once, which is the appropriate target if one \ITZ\ model underlies all of them, but this model reports only the descriptors consistent with every specimen simultaneously and drives the differences between them into the discrepancy term.
The pooled posterior is formulated in \cref{sec:posterior}.
The opposite choice treats each specimen in isolation, giving every one its own descriptor vector; this lets the geometries disagree, but it discards the fact that the specimens are extractions of one material and so says nothing about the \ITZ\ at a further unobserved location.
A hierarchical model, developed in \cref{sec:hierarchical}, sits between these extremes: each specimen keeps its own descriptors, but they are drawn from a common population whose mean and spread are inferred, so that the specimens inform one another and jointly characterize both the material and its spatial variability.

We adopt the hierarchical model as our primary inference target: the specimens plainly differ, and only a population model supplies the predictive descriptors of an unobserved aggregate that mesoscale propagation requires.
The pooled posterior remains useful throughout as a baseline, and is the only target for which cheaper alternatives exist that are well posed.
\cref{sec:point-estimate,sec:laplace} give two such alternatives that forgo sampling, a deterministic point estimate from a regularized least-squares fit and a Laplace approximation that fits a Gaussian to the posterior at its mode, both applied to the pooled posterior.
\cref{sec:sampling} instead characterizes the full posterior directly, by sampling, which is also how we treat the hierarchical model.

\subsection{The likelihood with model discrepancy}
\label{sec:likelihood}

We index the \(\nspec = 10\) digital specimens of \cref{sec:itz-specimen} by \(g = 1, \dots, \nspec\), and let \(\fwd_g(\params)\) collect the homogenized-model displacements at the sensors of specimen \(g\).
The homogenized model shares one geometry across all specimens, meaning only one \FEM-solve per likelihood evaluation.

If the homogenized model were exact, the data of specimen \(g\) would differ from \(\fwd_g(\params)\) by measurement noise alone.
However, it is not exact: reducing the resolved microstructure to a smooth axial stiffness gradient discards features that leave a systematic imprint on the surface displacements.
Treating this imprint as independent noise per sensor would misrepresent it, because the mismatch varies smoothly over the beam surface and is therefore correlated between nearby sensors.
We follow Kennedy and O'Hagan~\cite{Kennedy2001a} and represent the mismatch by an additive discrepancy field \(\delta(\vx)\), modelled a priori as a zero-mean Gaussian process over the beam surface,
\begin{equation}
  \delta \sim \GP{0}{\disckernel}, \qquad \disckernel(\vx, \vx') = \discvar \exp\!\left( -\frac{\norm{\vx - \vx'}^2}{2 \corrlength^2} \right),
  \label{eq:discrepancy-gp}
\end{equation}
with a squared-exponential covariance whose variance \(\discvar\) sets the scale of the systematic mismatch and whose correlation length \(\corrlength\) sets the distance over which it varies across the surface.

The observation model for specimen \(g\) superposes the model prediction, a realization of the discrepancy at its sensors, and measurement noise,
\begin{equation}
  \vy_g = \fwd_g(\params) + \discrepancy_g + \noisevec_g,
  \qquad
  \discrepancy_g \sim \normal{\vnull}{\corrmat_g},
  \quad
  \noisevec_g \sim \normal{\vnull}{\noisevar \mI},
  \label{eq:observation-model}
\end{equation}
where \(\corrmat_g\) is the discrepancy correlation matrix with entries \([\corrmat_g]_{pq} = \disckernel(\vx_p, \vx_q)\) evaluated at the sensor locations \(\vx_p\) of specimen \(g\), and the noise is independent across sensors with variance \(\noisevar\).
Each sensor records the three Cartesian components of the surface displacement; the discrepancy kernel acts on the spatial separation of the sensors and is shared across components, which are taken to be mutually independent, so that the full covariance is block diagonal with one identical \(\corrmat_g + \noisevar \mI\) block per component.

We are not interested in the discrepancy for its own sake, only in its effect on the inference of \(\params\).
Because \(\discrepancy_g\) and \(\noisevec_g\) are both Gaussian, we marginalize the discrepancy analytically, which leaves a Gaussian likelihood for each specimen,
\begin{equation}
  \probabilityof[\params, \hyperparams]{\vy_g} = \normal{\fwd_g(\params)}{\obscov_g},
  \qquad
  \obscov_g = \corrmat_g + \noisevar \mI.
  \label{eq:per-specimen-likelihood}
\end{equation}
The covariance \(\obscov_g\) superposes the correlated discrepancy and the diagonal noise, so the model error enters the likelihood as structured, spatially correlated uncertainty rather than being absorbed into an inflated noise level.
The three quantities \(\hyperparams = (\discvar, \noisevar, \corrlength)\) are not known beforehand and are inferred alongside \(\params\).

The ten specimens are independent extractions of the \ITZ\ microstructure (\cref{sec:itz-specimen}), so we treat their datasets as conditionally independent given the parameters, sharing the hyperparameters \(\hyperparams\) but each carrying its own discrepancy realization.
The joint likelihood is therefore the product over specimens,
\begin{equation}
  \probabilityof[\params, \hyperparams]{\vy_1, \dots, \vy_{\nspec}} = \prod_{g=1}^{\nspec} \normal*[\vy_g]{\fwd_g(\params)}{\obscov_g},
  \label{eq:joint-likelihood}
\end{equation}
with log-likelihood
\begin{equation}
  \log \probabilityof[\params, \hyperparams]{\vy_1, \dots, \vy_{\nspec}} = -\frac{1}{2} \sum_{g=1}^{\nspec} \left[ \transpose{\vr_g} \inverse{\obscov_g} \vr_g + \log\absolute{\obscov_g} \right] + \text{const},
  \qquad
  \vr_g = \vy_g - \fwd_g(\params),
  \label{eq:log-likelihood}
\end{equation}
where \(\vr_g\) is the residual between the data and the model prediction for specimen \(g\).

\subsection{Prior}
\label{sec:prior}

We complete the Bayesian model with priors on the descriptors \(\params\) and the hyperparameters \(\hyperparams\).
The descriptors are physically constrained: the modulus ratio \(\modratio\) lies in \((0, 1]\), the recovery length \(\lengthscale\) is positive, and the bulk-paste modulus \(\Einf\) is positive and bounded above by the stiffness of the anhydrous cement particle, the stiffest paste constituent at \SI{99}{\giga\pascal} (\cref{sec:itz-specimen}).
We enforce these ranges by reparameterizing each descriptor onto an unconstrained scale --- a logit transform for the bounded \(\modratio\) and \(\Einf\), and a logarithm for the positive \(\lengthscale\) --- and likewise map the positive hyperparameters \(\discvar\), \(\noisevar\), and \(\corrlength\) to their logarithms.
We write this componentwise reparameterization of the descriptors as \(\descl = \reparam(\params)\), with inverse \(\params(\descl) = \inverse{\reparam}(\descl)\), and collect the six transformed variables into a single latent vector \(\latent = (\descl, \log \hyperparams)\), whose first three components \((\xi_1, \xi_2, \xi_3)\) form the descriptor block \(\descl\) and whose last three \((\xi_4, \xi_5, \xi_6)\) are the componentwise logarithm of the discrepancy hyperparameters \(\hyperparams\); we assign each component an independent Gaussian prior.
We fix each Gaussian prior by propagating a physical-space mean and standard deviation through its transform, so that the prior is interpretable on the scale of the quantity it constrains.

\begin{figure}[tbp]
  \centering
  \includegraphics[width=\textwidth]{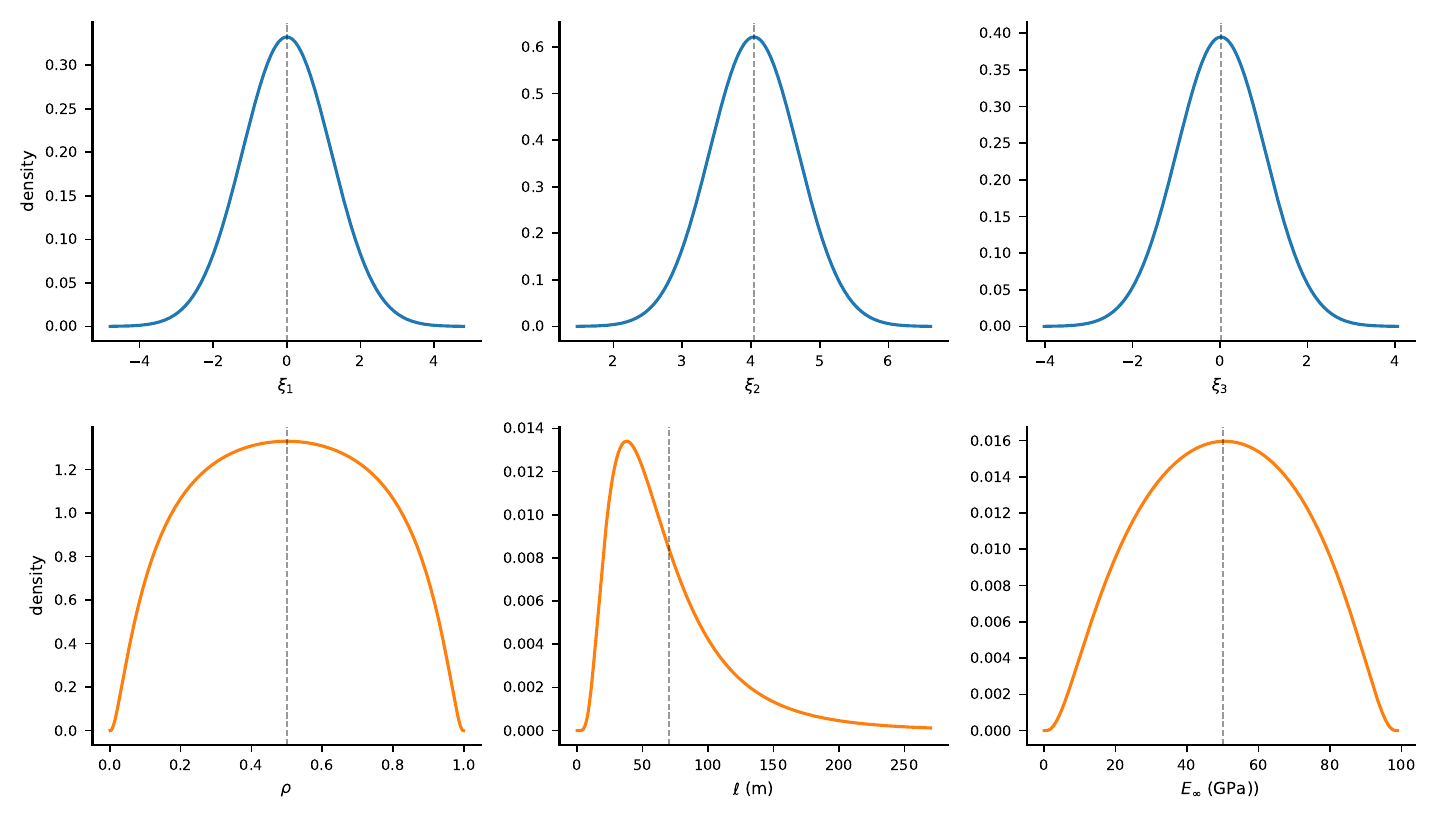}
  \caption{Prior distributions of the three \ITZ\ descriptors \(\params = (\modratio, \lengthscale, \Einf)\). The top row shows the Gaussian priors on the unconstrained latent variables \(\xi_1, \xi_2, \xi_3\); the bottom row shows the corresponding densities pushed forward to physical space, where the logit and logarithm transforms confine \(\modratio\) to \((0, 1)\) and \(\Einf\) to \((0, \SI{99}{\giga\pascal})\) and confine the recovery-length prior on \(\lengthscale\) to $\R^+$. Dashed lines mark the prior mean in each space.}
  \label{fig:prior-descriptors}
\end{figure}

\begin{figure}[tbp]
  \centering
  \includegraphics[width=\textwidth]{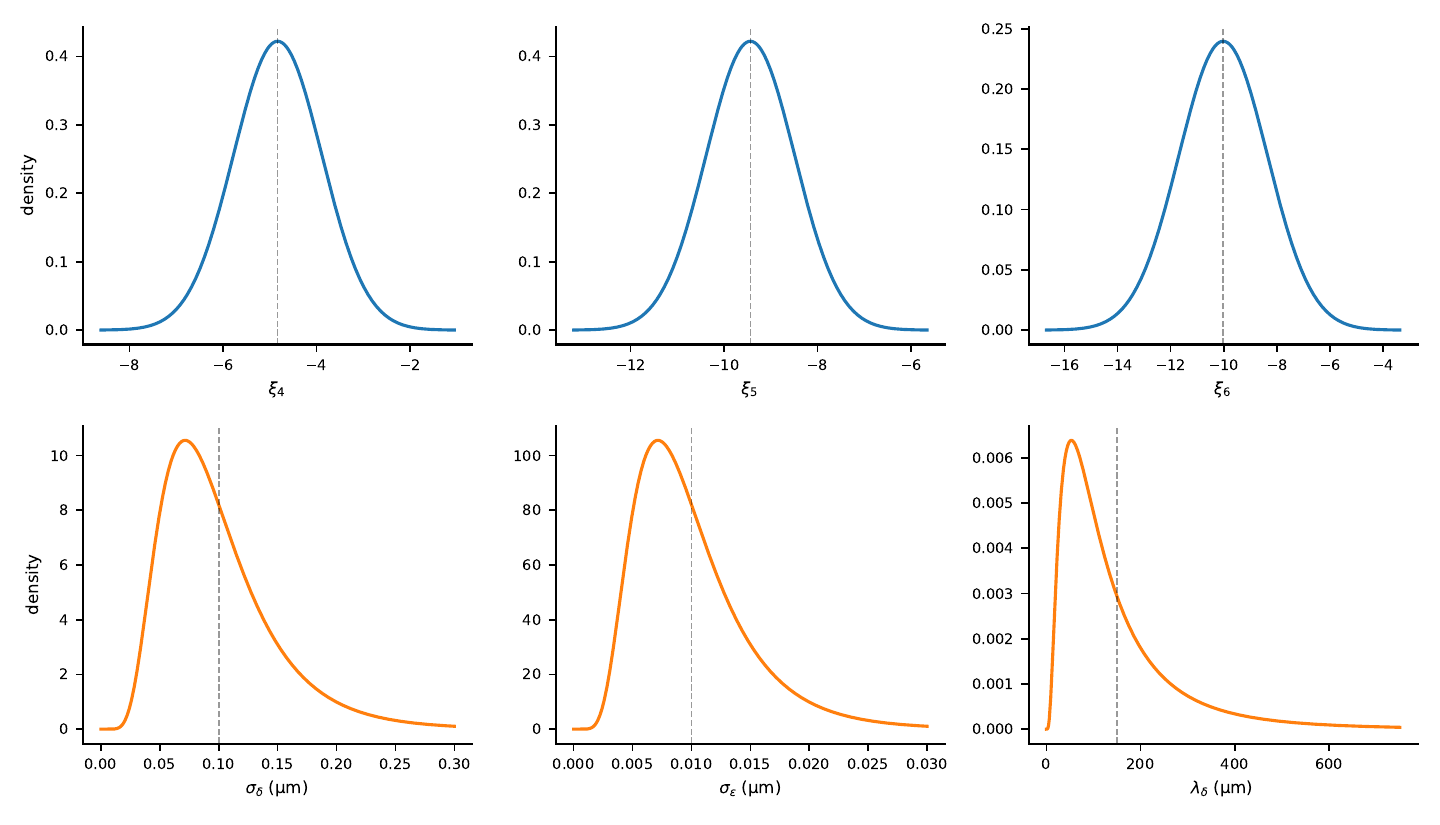}
  \caption{Prior distributions of the three discrepancy hyperparameters \(\hyperparams\). The top row shows the Gaussian priors on the latent variables \(\xi_4, \xi_5, \xi_6\); the bottom row shows the log-normal physical-space priors on the discrepancy amplitude \(\sigma_{\delta}\), the noise amplitude \(\sigma_{\epsilon}\), and the correlation length \(\corrlength\). Dashed lines mark the prior mean in each space.}
  \label{fig:prior-hyperparams}
\end{figure}

\Cref{fig:prior-descriptors,fig:prior-hyperparams} show the resulting marginals in both the latent space, where they are Gaussian by construction, and the physical space, where the transforms render them bounded and skewed.
For the descriptors we centre the modulus ratio at \(\modratio = 0.5\) with standard deviation \num{0.3}, the recovery length at \(\lengthscale = \SI{70}{\micro\meter}\) with standard deviation \SI{50}{\micro\meter}, and the bulk modulus at \(\Einf = \SI{50}{\giga\pascal}\) with standard deviation \SI{25}{\giga\pascal}; these spans are deliberately broad, covering the physically plausible range without favouring a weakened or an intact interface.

For the hyperparameters we centre the discrepancy amplitude \(\sigma_{\delta}\) at \SI{0.1}{\micro\meter} with standard deviation \SI{0.05}{\micro\meter}, the noise amplitude \(\sigma_{\epsilon}\) at \SI{0.01}{\micro\meter} with standard deviation \SI{0.005}{\micro\meter} --- centred on the noise level used to generate the measurements --- and the correlation length \(\corrlength\) at \SI{150}{\micro\meter} with standard deviation \SI{150}{\micro\meter}.
The discrepancy amplitude is thus kept small, of the order of the measured displacements, while the broad correlation-length prior leaves the scale over which the model error varies across the surface to be fixed by the data; in this sense both hyperparameter priors are only weakly informative.
The prior on the noise amplitude is informative, as we encode our knowledge about the measurement process, but still allowing for some uncertainty about the exact noise level.

\subsection{Pooled posterior}
\label{sec:posterior}

Given the likelihood and priors, the posterior distribution for the pooled scenario follows from Bayes' theorem:
\begin{equation}
  \probabilityof[\vy_1, \dots, \vy_{\nspec}]{\params, \hyperparams} \propto \probabilityof[\params, \hyperparams]{\vy_1, \dots, \vy_{\nspec}} \, \probabilityof{\params} \, \probabilityof{\hyperparams}.
  \label{eq:posterior}
\end{equation}
This joint posterior is the target of the inference; its marginal over \(\params\) is the quantity that answers our question, summarizing the \ITZ\ descriptors and their uncertainty once the discrepancy hyperparameters have been integrated out.
The pooled posterior of \cref{eq:posterior} forces a single descriptor vector on every specimen.

\subsection{A deterministic point estimate}
\label{sec:point-estimate}

Calibrating a model by regularized maximum-likelihood fitting is the conventional approach to inverse problems of this kind across many fields, and we adopt such a point estimate as the reference against which to measure what the Bayesian treatment adds.
For this estimate we replace the likelihood by a standard Gaussian likelihood with fixed measurement-noise variance, the special case \(\discvar = 0\) of \cref{eq:per-specimen-likelihood},
\begin{equation}
  \probabilityof[\params]{\vy_g} = \normal*[\vy_g]{\fwd_g(\params)}{\noisevar \mI}.
  \label{eq:standard-likelihood}
\end{equation}
We omit the discrepancy because it is probabilistic in nature and not well represented by a point estimate.
Pooling the specimens, the joint log-likelihood is then a plain sum of squares,
\begin{equation}
  \log \probabilityof[\params]{\vy_1, \dots, \vy_{\nspec}} = -\frac{1}{2\noisevar} \sum_{g=1}^{\nspec} \norm{\vr_g}^2 + \text{const},
  \qquad
  \vr_g = \vy_g - \fwd_g(\params).
  \label{eq:standard-loglik}
\end{equation}

The inverse problem is ill-posed, so this likelihood alone has no usable maximizer; the reparameterization of \cref{sec:prior} bounds the descriptors but does not remove the ill-posedness.
The Gaussian prior supplies the missing regularization and ensures a comparison on equal footing with the Bayesian treatment.
With \(\discvar\) and \(\noisevar\) both fixed rather than inferred, the only unknown left is the descriptor latent vector \(\descl\) of \cref{sec:prior}, and we take the point estimate to be the maximum a posteriori (\MAP) point of the pooled posterior over \(\descl\) alone,
\begin{equation}
  \descl^{\star} = \argmaxof[\descl]{\log \probabilityof[\params(\descl)]{\vy_1, \dots, \vy_{\nspec}} + \log \probabilityof{\descl}}.
  \label{eq:map}
\end{equation}
With a Gaussian likelihood and a Gaussian prior, this is a Tikhonov-regularized nonlinear least-squares estimate, with the likelihood as the data misfit and the prior a quadratic penalty.
We compute it for the pooled posterior alone, locating the \MAP\ in the latent space of \cref{sec:prior}, where the Gaussian prior, and with it the regularization, is defined.

\subsection{The Laplace approximation}
\label{sec:laplace}

The point estimate of \cref{sec:point-estimate} discards both the discrepancy and the uncertainty in the descriptors, reporting a single best-fitting latent vector.
We recover a posterior, with its uncertainty, while remaining in the realm of deterministic optimization through a Laplace approximation: a Gaussian fitted to the pooled posterior of \cref{eq:posterior} at its mode.
Unlike the point estimate, the Laplace approximation retains the full discrepancy likelihood of \cref{eq:per-specimen-likelihood} and so inhabits the same probabilistic model as the sampled posterior, over the full latent vector \(\latent\) of descriptors and hyperparameters.
It centres the Gaussian at the posterior mode and sets its covariance from the curvature of the log-posterior there,
\begin{equation}
  \begin{split}
    \probabilityof[\vy_1, \dots, \vy_{\nspec}]{\latent} &\approx \normal*[\latent]{\latentmap}{\inverse{\mH}}, \\
    \latentmap &= \argmaxof[\latent]{\log \probabilityof[\vy_1, \dots, \vy_{\nspec}]{\latent}},\\
    \mH &= -\left. \nabla_{\latent}^2 \log \probabilityof[\vy_1, \dots, \vy_{\nspec}]{\latent} \right|_{\latent = \latentmap},
  \end{split}
  \label{eq:laplace}
\end{equation}
where the mode \(\latentmap\) is found by the same maximization as the point estimate of \cref{eq:map}, now applied to the full posterior rather than its discrepancy-free special case, and \(\mH\) is the Hessian of the negative log-posterior at the mode, whose inverse gives the approximate covariance.
The approximation is exact when the posterior is Gaussian, which here it is not: the forward map \(\fwd_g\) depends nonlinearly on the descriptors, so \cref{eq:laplace} amounts to linearizing \(\fwd_g\) about \(\latentmap\), and is accurate only in the vicinity of the posterior mode.
We apply this approximation only to the pooled posterior.

Evaluating \(\mH\) dominates the cost of the approximation.
Since the finite-element library differentiates the solve of \cref{eq:linear-system} in reverse mode only, we form \(\mH\) by central differences of the adjoint gradient along the three descriptors \(\descl\); the hyperparameters enter \(\obscov_g\) in closed form and need no additional solves.
This costs \num{12} linear solves, each with a different stiffness matrix, and amplifies errors in the gradient by \(1/h\) for the step \(h\).
With forward mode available, exact Hessian--vector products would reduce this to eight solves sharing a single factorization of \(\mK\), whereas differencing the log-posterior itself would require \num{13} forward solves with an error growing as \(1/h^2\).

\subsection{A hierarchical model for aggregate-to-aggregate variability}
\label{sec:hierarchical}

The opposite extreme of the pooled posterior treats each specimen as its own inference problem, with its own descriptors \(\params_g\) and hyperparameters \(\hyperparams_g\), and its posterior following from the per-specimen likelihood of \cref{eq:per-specimen-likelihood} and the same priors,
\begin{equation}
  \probabilityof[\vy_g]{\params_g, \hyperparams_g} \propto \normal*[\vy_g]{\fwd_g(\params_g)}{\obscov_g} \, \probabilityof{\params_g} \, \probabilityof{\hyperparams_g},
  \qquad g = 1, \dots, \nspec.
  \label{eq:per-specimen-posterior}
\end{equation}
This lets the geometries disagree and exposes the specimen-to-specimen variability that pooling averages into its discrepancy term, but it goes too far in the other direction: it treats the \(\nspec\) specimens as unrelated, though they are extractions of one material, and it leaves the descriptors of an unobserved ITZ --- the quantity the mesoscale propagation requires --- undefined.
A hierarchical model occupies the middle ground between these two extremes.
The specimens differ, but they are draws from a common population, so each specimen carries information about the others and about a further aggregate that has not been observed.

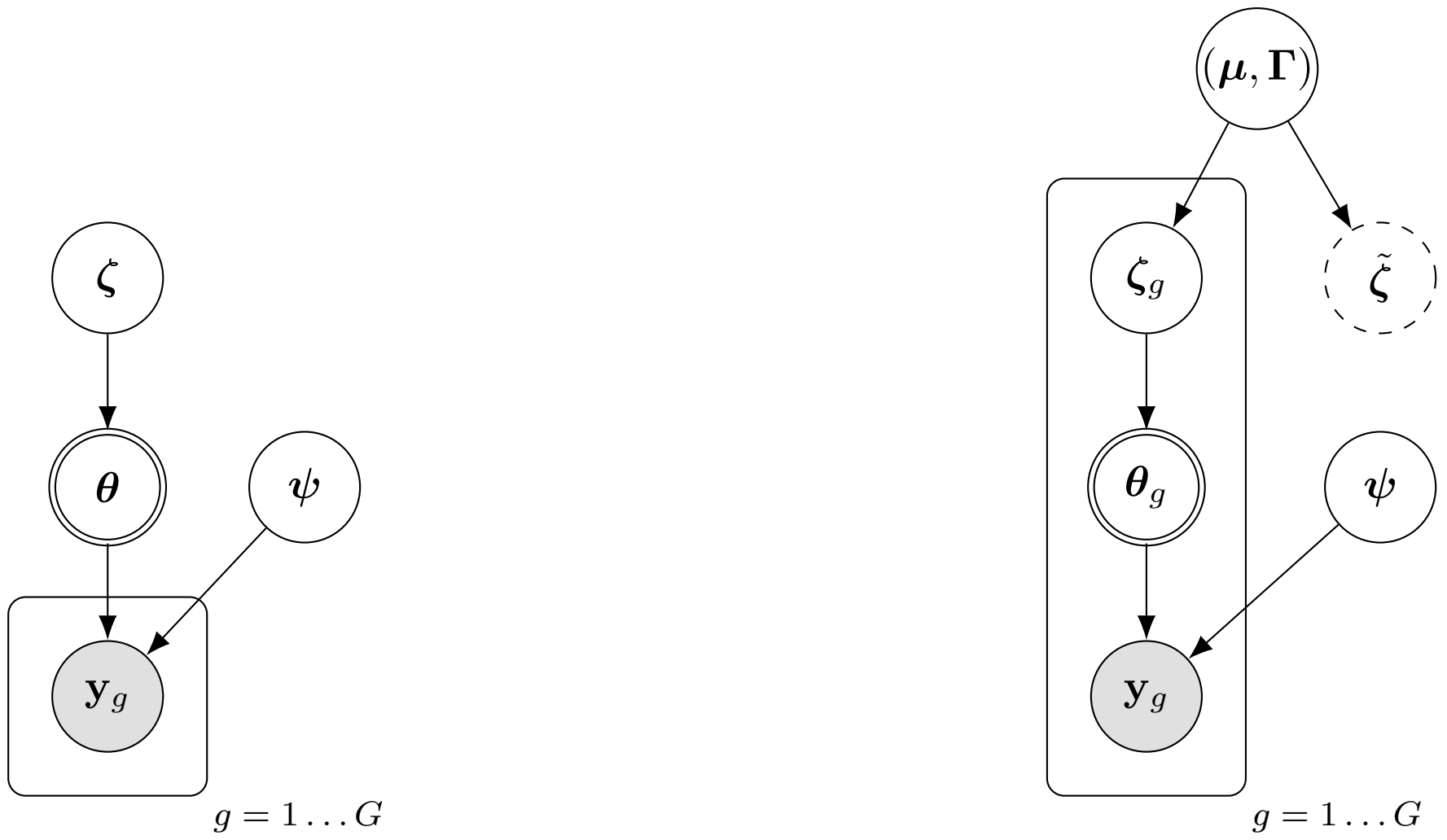
\begin{figure}[tbp]
  \centering
  \subcaptionbox{Pooled, \cref{eq:posterior}\label{fig:dag-pooled}}[0.47\textwidth]{%
    \begin{tikzpicture}[
        baseline=(y),
        var/.style={circle, draw, minimum size=9mm, inner sep=0pt, font=\small},
        obs/.style={var, fill=black!12},
        det/.style={var, double, double distance=1pt},
        pl/.style={draw, rounded corners, inner sep=3.5mm},
        ar/.style={-{Latex[length=2mm]}},
      ]
      \node[var] (zeta)  at (0,4.0) {\(\descl\)};
      \node[det] (theta) at (0,2.3) {\(\params\)};
      \node[var] (psi)   at (1.6,2.3) {\(\hyperparams\)};
      \node[obs] (y)     at (0,0.6) {\(\vy_g\)};
      \node[pl, fit=(y), label={[font=\scriptsize, inner sep=1pt]south east:\(g = 1 \dots \nspec\)}] (plate) {};
      \draw[ar] (zeta) -- (theta);
      \draw[ar] (theta) -- (y);
      \draw[ar] (psi) -- (y);
    \end{tikzpicture}}%
  \hfill
  \subcaptionbox{Hierarchical, \cref{eq:hier-model}\label{fig:dag-hierarchical}}[0.47\textwidth]{%
    \begin{tikzpicture}[
        baseline=(y),
        var/.style={circle, draw, minimum size=9mm, inner sep=0pt, font=\small},
        obs/.style={var, fill=black!12},
        det/.style={var, double, double distance=1pt},
        pred/.style={var, dashed},
        pl/.style={draw, rounded corners, inner sep=3.5mm},
        ar/.style={-{Latex[length=2mm]}},
      ]
      \node[var]  (pop)   at (0.9,5.7) {\(\poppar\)};
      \node[var]  (zeta)  at (0,4.0) {\(\descl_g\)};
      \node[det]  (theta) at (0,2.3) {\(\params_g\)};
      \node[var]  (psi)   at (1.9,2.3) {\(\hyperparams\)};
      \node[obs]  (y)     at (0,0.6) {\(\vy_g\)};
      \node[pl, fit=(zeta)(theta)(y), label={[font=\scriptsize, inner sep=1pt]south east:\(g = 1 \dots \nspec\)}] (plate) {};
      \node[pred] (new)   at (1.9,4.0) {\(\newdesc\)};
      \draw[ar] (pop) -- (zeta);
      \draw[ar] (pop) -- (new);
      \draw[ar] (zeta) -- (theta);
      \draw[ar] (theta) -- (y);
      \draw[ar] (psi) -- (y);
    \end{tikzpicture}}
  \caption{
    The pooled and hierarchical models as directed graphs.
    Shaded nodes are observed, and the plate repeats its contents over the \(\nspec\) specimens.
    Double circles are deterministic: the physical descriptors \(\params = \inverse{\reparam}(\descl)\) are a fixed function of the latent descriptors of \cref{sec:prior} and carry no distribution of their own.
    (a) One descriptor vector \(\descl\) explains every specimen.
    (b) Each specimen keeps its own descriptors \(\descl_g\), drawn from a population \(\poppar\); the dashed node \(\newdesc\) is an unobserved \ITZ, predicted from the same population.
  }
  \label{fig:dags}
\end{figure}

\Cref{fig:dags} contrasts the pooled and hierarchical models structurally, as directed graphs.
In the pooled model of \cref{fig:dag-pooled}, a single node \(\descl\) feeds every specimen, so all \(\nspec\) datasets are evidence about the same quantity and the posterior contracts as though the specimens were repeated measurements of one object.
In the hierarchical model of \cref{fig:dag-hierarchical}, the descriptors are specific to the specimen but descriptors for all specimen share a parent, the population parameters \(\poppar\).
The dashed node \(\newdesc\), attached to the same parent but outside the plate, is the descriptor of the \ITZ\ of an aggregate we have not observed, which can be used for the mesoscale propagation.

Formally, we place the hierarchy on the descriptor block \(\descl\) of the latent vector \(\latent = (\descl, \log \hyperparams)\) of \cref{sec:prior} alone, leaving the discrepancy hyperparameters \(\hyperparams\) untouched.
The hyperparameters remain shared across specimens as in the pooled model, as the model discrepancy is a property of the forward model and the measurement process, not of the specimens themselves.
Working on the latent scale is convenient, because the descriptors are unbounded there, so that a Gaussian population distribution respects the physical constraints of \cref{sec:prior} automatically once pushed through the transforms.
The model is
\begin{equation}
  \begin{split}
    \popmean, \popcov &\sim \probabilityof{\popmean} \, \probabilityof{\popcov}, \\
    \descl_g \mid \popmean, \popcov &\sim \normal{\popmean}{\popcov}, \qquad g = 1, \dots, \nspec, \\
    \vy_g \mid \descl_g, \hyperparams &\sim \normal{\fwd_g(\params(\descl_g))}{\obscov_g},
  \end{split}
  \label{eq:hier-model}
\end{equation}
with \(\obscov_g = \corrmat_g + \noisevar \mI\) exactly as in \cref{eq:per-specimen-likelihood}.
The population mean \(\popmean\) is the descriptor vector of the material, and the population covariance \(\popcov\) is the variability of that descriptor vector between \ITZ s of different aggregates.
We take \(\popcov = \diagof{\popsd_1^2, \popsd_2^2, \popsd_3^2}\), because we have insufficient information to identify a free correlation structure, and place a half-normal prior on the scales \(\popsd_j\), sampled as \(\log \popsd_j\).
The half-normal is preferred over an inverse-gamma prior on the variances, which is poorly behaved near \(\popsd_j = 0\)~\cite{Gelman2006}.
For \(\popmean\) and \(\hyperparams\) we reuse the priors of \cref{sec:prior} unchanged.

The pooled and per-specimen targets are the two limits of \cref{eq:hier-model}.
Sending \(\popcov \to \vnull\) collapses every \(\descl_g\) onto \(\popmean\) and recovers the pooled posterior of \cref{eq:posterior}, while \(\popcov \to \infty\) severs the specimens and recovers the \(\nspec\) independent posteriors of \cref{eq:per-specimen-posterior}.
Inferring \(\popcov\) rather than fixing it at either extreme is what lets the data decide how much the specimens should inform one another.

Two objects follow from \cref{eq:hier-model}; the marginal posterior \(\probabilityof[\alldata]{\popmean}\) estimates the material's typical \ITZ, the centre of the population, as a quantity distinct from its aggregate-to-aggregate spread \(\popcov\).
The posterior predictive
\begin{equation}
  \probabilityof[\alldata]{\newdesc}
  = \int \normal*[\newdesc]{\popmean}{\popcov} \, \probabilityof[\alldata]{\popmean, \popcov} \dif{\popmean} \dif{\popcov}
  \label{eq:hier-predictive}
\end{equation}
is the distribution of the descriptors \(\newdesc\) of an unobserved \ITZ\ not among the \(\nspec\).

Both objects are computed from the joint posterior implied by \cref{eq:hier-model}, which by Bayes' theorem is, analogously to \cref{eq:posterior},
\begin{equation}
  \probabilityof[\alldata]{\descl_{1:\nspec}, \popmean, \popcov, \hyperparams}
  \propto
  \left[ \prod_{g=1}^{\nspec} \normal*[\vy_g]{\fwd_g(\params(\descl_g))}{\obscov_g} \, \normal*[\descl_g]{\popmean}{\popcov} \right]
  \probabilityof{\popmean} \, \probabilityof{\popcov} \, \probabilityof{\hyperparams}.
  \label{eq:hier-posterior}
\end{equation}
This joint posterior, sampled over \((\descl_{1:\nspec}, \popmean, \log \popsds, \log \hyperparams)\), has dimension \(3\nspec + 9 = \num{39}\) for \(\nspec = 10\).
While the specimen descriptors \(\descl_g\) are not the primary target of the inference, we must first sample the joint of \cref{eq:hier-posterior} to then obtain the marginal \(\probabilityof[\alldata]{\popmean, \popcov}\).
The joint posterior has no closed form; \cref{sec:sampling} describes how we sample it, alongside the pooled posterior.

The point estimate and Laplace approximation of \cref{sec:point-estimate,sec:laplace} do not extend to this hierarchical model.
Both rely on a mode of the target density, and the hierarchical posterior of \cref{eq:hier-model} has none: its centred joint density is unbounded along the ray where every specimen sits on the population mean and the population scales shrink to zero.
The point estimate has no maximizer to converge to and the Laplace approximation has nothing to build its Gaussian around.
This gap is not just technical -- collapsing a population model to a single point would discard the aggregate-to-aggregate spread \(\popcov\) that the hierarchical model exists to quantify in the first place.

\subsection{Sampling the full posterior}
\label{sec:sampling}

\begin{figure}[tbp]
  \centering
  \includegraphics[width=\textwidth]{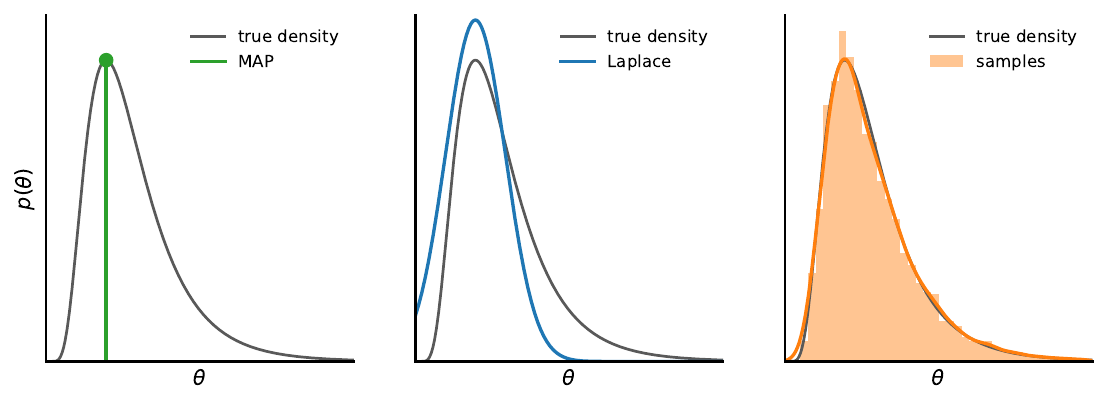}
  \caption{
    Schematic illustration of the three characterizations of \cref{sec:point-estimate,sec:laplace,sec:sampling}, on a log-normal density.
    Left: the point estimate, retaining only the mode.
    Centre: the Laplace approximation.
    Right: the full density approximated by sampling.
  }
  \label{fig:inference-schematic}
\end{figure}

The full posterior has no closed form, so we characterize it directly by sampling.
\Cref{fig:inference-schematic} illustrates the three characterizations of \cref{sec:point-estimate,sec:laplace,sec:sampling} schematically, on a generic one-dimensional target rather than the actual \ITZ\ posterior.
The point estimate keeps only the mode; the Laplace approximation adds a Gaussian shape around it, set by the local curvature; sampling recovers the full density, skew included.
We use the \BPS, a piecewise-deterministic Markov process (\PDMP), within the surrogate-assisted sampling framework developed for \PDE-constrained inverse problems of this kind~\cite{Riccius2026b} to sample from the posterior.
The \BPS\ traverses the posterior along a piecewise-linear trajectory that reflects its velocity at the events of a Poisson process, also called the bounce process.
At events of an additional refreshment Poisson process with constant rate \(\refreshrate\), the velocity is randomized to ensure that the sampler can explore the entire parameter space.
The rate of the bounce process is the positive part of the directional derivative of the potential -- the negative log-posterior -- along the current velocity.
Simulating the sampler therefore requires gradients of the potential, not merely its values.

The forward model is implemented in the differentiable finite-element library JAX-FEM~\cite{Xue2022}, which efficiently supplies exact gradients of the log-likelihood with respect to \(\params\) through reverse-mode differentiation of the solver.
The gradients with respect to \(\hyperparams\) follow in closed form from \cref{eq:log-likelihood}.
The sampler operates on the unconstrained, reparameterized variables.

We run the standard version of this sampler for the pooled posterior of \cref{eq:posterior}.
The trajectory starts from the latent-space origin \(\latent = \vnull\) with an initial velocity drawn uniformly on the unit sphere, and is simulated until \(T = \num{10000}\) in the sampler's intrinsic time, with the velocity refreshed at rate \(\refreshrate = 0.1\).
We discretize the resulting path at \num{20000} evenly spaced times, transform the samples back to the original parameter space, and report them as the posterior draws.

Following the surrogate-assisted construction of~\cite{Riccius2026b}, the reflection events are generated not from the exact potential but from a cheaper upper bound built on a surrogate, with a subsequent thinning step that ensures the sampler remains exact.
The surrogate is a Gaussian-process (\GPabbr) model of the potential, fitted once before sampling.
The affine reparameterization of \cref{sec:prior} whitens the posterior around the \MAP\ point, so the \GPabbr\ need only represent the departure of the log-posterior from its Gaussian base.
We train the \GPabbr\ model on \num{1000} points drawn from this base, evaluating the potential at each, and fit a squared-exponential kernel with one length scale per latent coordinate by maximizing the marginal likelihood with L-BFGS, retaining the best of \num{50} restarts.
During sampling the gradient of the \GPabbr\ posterior mean supplies the proposal rate, while the thinning step against the exact potential guarantees that the draws target the true posterior.
The sampler demanded a total of 7147 likelihood gradient evaluations, where \SI{78.9}{\percent} of proposed events were accepted.

The hierarchical posterior of \cref{eq:hier-model} is sampled separately.
Its \num{39}-dimensional joint density couples the ten specimens through the shared population parameters and hyperparameters, so the single-surrogate construction above no longer applies directly.
We instead exploit the factorized structure of the target with a local \BPS, in which each factor of the potential carries its own Poisson clock and a bounce reflects only the coordinates that factor reads.
Because each candidate event involves a single factor, the per-event cost stays at one finite-element solve despite the ten forward models in the target.
\Cref{app:local-bps} sets out the factor graph, the per-specimen quadratic surrogates, the mass preconditioner, and the adaptive thinning that together make this sampler practical, along with the measured cost of a run.

\section{Forward propagation of the posterior}
\label{sec:forward-uq}

After inference, we push the descriptor posterior forward through a separate model of a \RVE populated with several aggregates, and report the induced distribution of mesoscale quantities of interest.
The map from descriptors to these quantities is nonlinear and available only through a mesoscale finite-element solve, so we approximate the pushforward by direct \MC:
we draw descriptor samples from the posterior, evaluate the \RVE\ model on each, and collect the outputs.
We carry out this propagation for both inference targets of \cref{sec:inference}.
The pooled posterior of \cref{eq:posterior} and the hierarchical predictive of \cref{eq:hier-predictive} differ in how the descriptors are distributed across the aggregates of the \RVE, and we treat the two cases separately below.

\subsection{The mesoscale RVE model}
\label{sec:rve-model}

The \RVE\ is a square two-dimensional cell \(\rvedomain\) of side length \(\rvesize = \SI{2}{\milli\meter}\) containing \(\naggr = \num{10}\) circular aggregates of prescribed centres and radii.
It is governed by the same small-strain isotropic linear elasticity as the forward model of \cref{sec:forward-model}, the equilibrium and constitutive relations of \crefrange{eq:equilibrium}{eq:constitutive}, now solved under plane-strain conditions and with the shared Poisson's ratio \(\poisson = \num{0.18}\).
We subject the \RVE\ to uniaxial tension along the \(x\)-axis: the axial macroscopic strain is held at \(\strain^{M}_{xx} = \num{0.01}\), while the transverse and shear components of the macroscopic stress are required to vanish, \ie \(\stress^{M}_{yy} = \stress^{M}_{xy} = 0\).
The displacement is represented as the affine part \(\macrostrains\,\vx\) plus a periodic fluctuation enforced through periodic boundary conditions on \(\rvedomain\), and the remaining components of \(\macrostrains\) follow from the stress-free conditions.
We discretize the \RVE\ with linear triangular elements and solve for the periodic fluctuation field.

\begin{figure}[tbp]
  \centering
  \subcaptionbox{Shared descriptor\label{fig:rve-efield}}{%
    \includegraphics[width=0.49\textwidth]{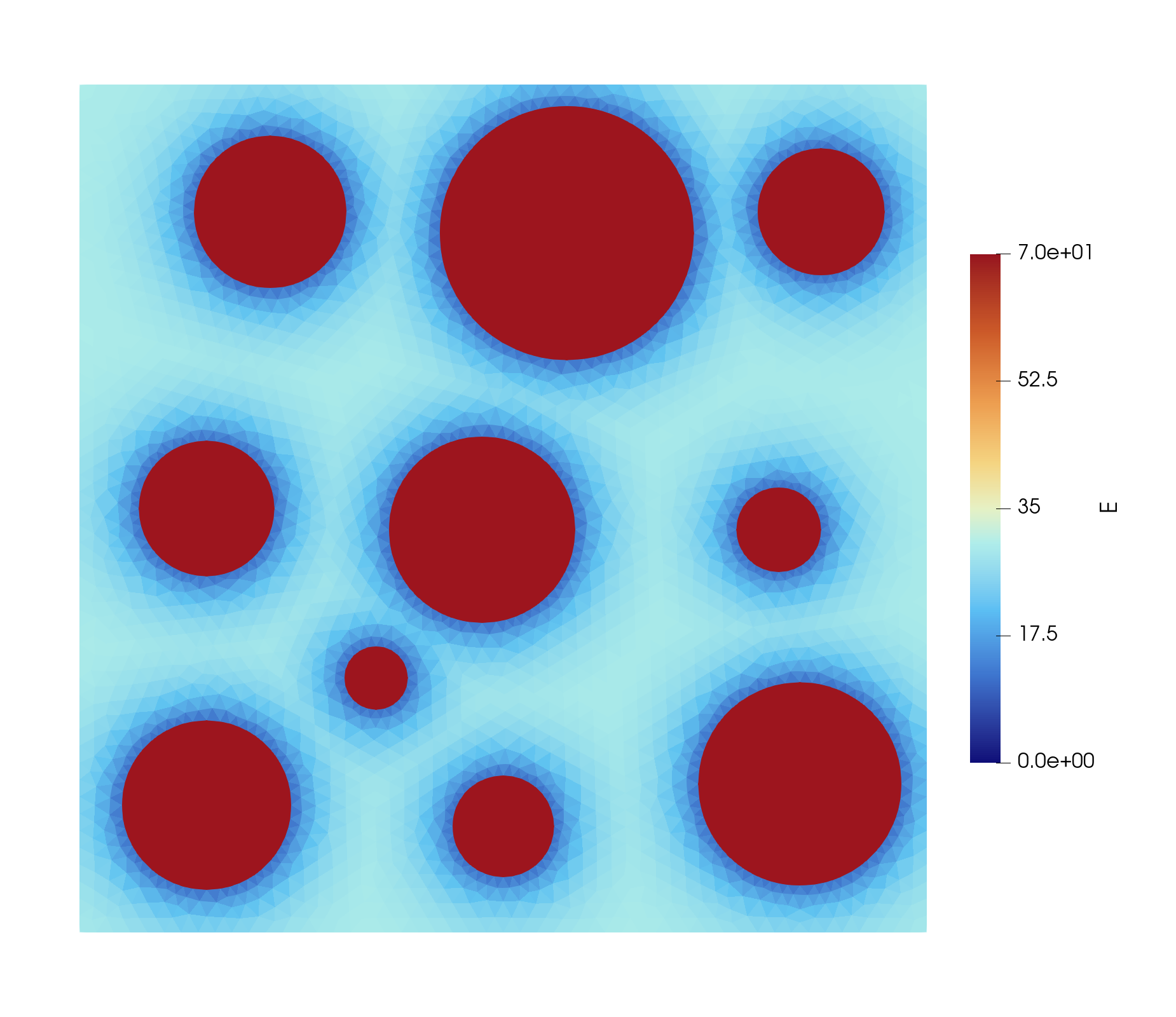}}%
  \subcaptionbox{Per-aggregate descriptor\label{fig:rve-efield-per-agg}}{%
    \includegraphics[width=0.49\textwidth]{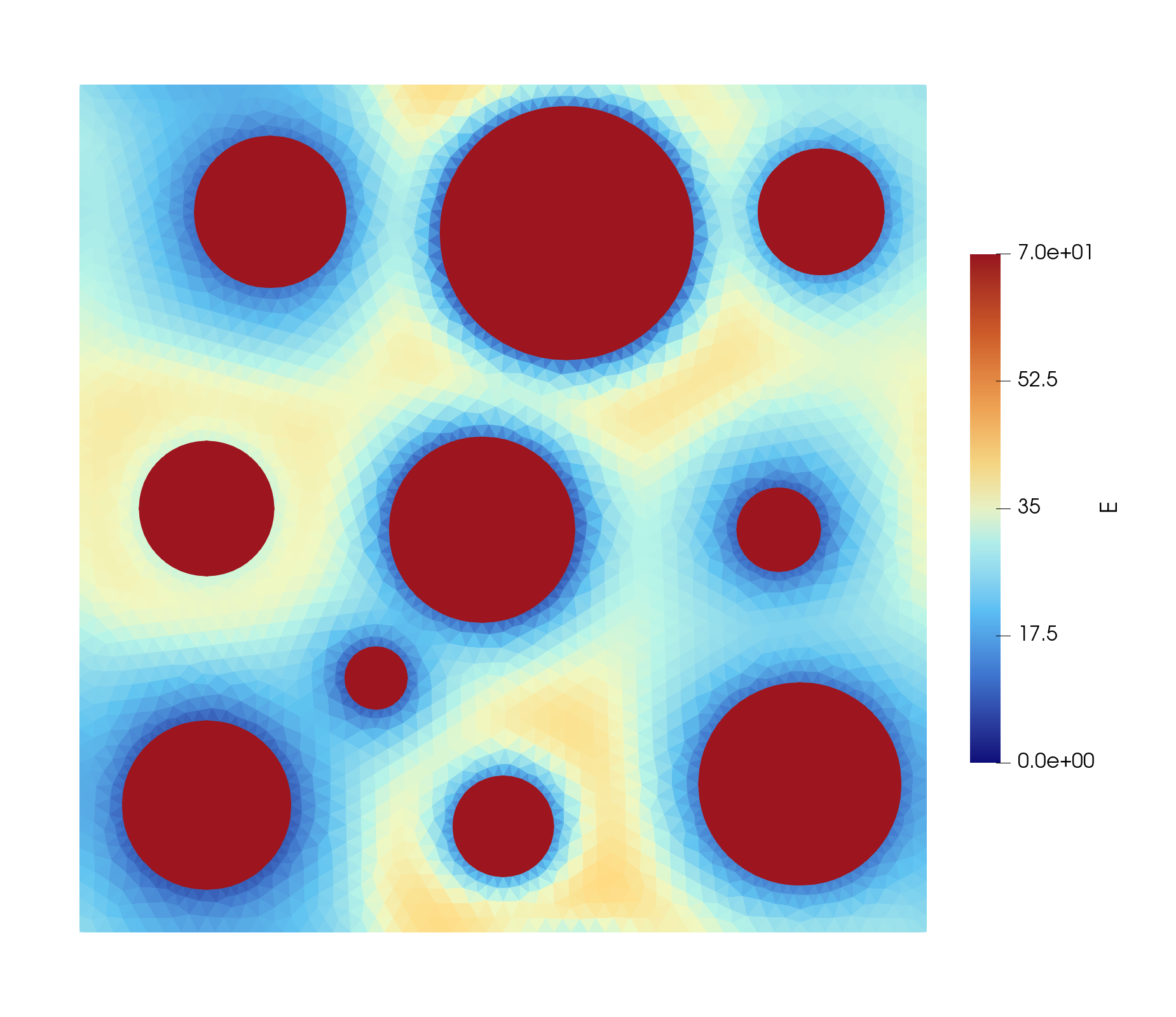}}%
  \caption{
    Young's modulus field of the \RVE\ for one \MC\ iteration.
    (a) Under the shared propagation, all aggregates share one descriptor, so the paste recovers to a single bulk modulus \(\Einf\).
    (b) Under the per-aggregate propagation each aggregate carries its own descriptor; neighbouring aggregates relax to visibly different bulk moduli.
  }
  \label{fig:rve-efields}
\end{figure}

The aggregates carry the constant aggregate modulus \(\Eagg = \SI{70}{\giga\pascal}\).
The surrounding paste carries an \ITZ\ stiffness gradient of the same form as the homogenized model of \cref{sec:homogenized-model}: the modulus recovers exponentially from a reduced near-interface value toward the bulk paste, but now measured from the aggregate surfaces rather than along a single beam axis.
Writing \(\distfield(\vx)\) for the distance from a point \(\vx\) in the paste to the surface of the nearest aggregate, the matrix modulus is
\begin{equation}
  \modulus(\vx) = \Einf \left[ 1 - (1 - \modratio)\, \exp\!\left( -\frac{\distfield(\vx)}{\lengthscale} \right) \right],
  \label{eq:rve-recovery-field}
\end{equation}
which is \cref{eq:recovery-field} with the axial offset \(\axial - \zitz\) replaced by the distance \(\distfield(\vx)\).
For a given \RVE, a draw of \(\params\) fully determines the modulus field, an example of such a field is given in \Cref{fig:rve-efield}.

From each solve we extract three quantities of interest that summarize the mesoscale response.
The first is the axial component \(\macrostressaxial\) of the volume-averaged stress \(\macrostresses = \volavg{\stresses}\), whose ratio to \(\strain^{M}_{xx}\) is the macroscopic uniaxial stiffness.
We also record the local extremes over the \RVE, the maximum von Mises stress \(\maxvonmises\) and the maximum principal strain \(\maxprincipal\), quantities that are relevant when considering onset of non-linearity.
We collect these three quantities into the output vector \(\qoi = (\macrostressaxial,\, \maxvonmises,\, \maxprincipal)^{\top}\).
Since a draw of \(\params\) fixes the modulus field of \cref{eq:rve-recovery-field}, and hence the entire \RVE\ response, this extraction defines the forward map
\begin{equation}
  \qoi = \rvemap(\params),
  \label{eq:rve-forward-map}
\end{equation}
from the \ITZ\ descriptors to the quantities of interest.

\subsection{Propagation of the pooled posterior}
\label{sec:propagate-linear}

For the pooled posterior of \cref{eq:posterior} a single descriptor vector \(\params\) governs the \ITZ\ around every aggregate, and the pushforward is obtained directly:
for each draw \(\params^{(i)}\), we build the modulus field of \cref{eq:rve-recovery-field}, solve the \RVE, and record the quantities \(\qoi^{(i)}\).
Their empirical distribution over the draws approximates the pushforward of the pooled posterior.
The direct \MC\ pushforward is exact in the large-sample limit but costs one \RVE\ solve per draw.
For the pooled posterior the Laplace approximation of \cref{sec:laplace} opens up a cheaper route: propagating its Gaussian through a linearization of the forward map --- a first-order second-moment estimate.
The Laplace approximation is the Gaussian \(\normal{\latentmap}{\inverse{\mH}}\) over the latent variables \(\latent\), and the quantities of interest depend on them through the descriptors, \(\qoi = \rvemap(\params(\latent))\), composing the reparameterization of \cref{sec:prior} with the \RVE\ map of \cref{eq:rve-forward-map}.
Linearizing this composition about the mode \(\latentmap\) propagates the Gaussian in closed form,
\begin{equation}
  \qoi \approx \normal*[\qoi]{\rvemap(\params(\latentmap))}{\mJ \inverse{\mH} \transpose{\mJ}},
  \qquad
  \mJ = \left. \nabla_{\latent}\, \rvemap(\params(\latent)) \right|_{\latent = \latentmap},
  \label{eq:linearized-pushforward}
\end{equation}
with the forward map at the mode for the mean.
The Laplace covariance is rescaled by the Jacobian \(\mJ\), which we compute via finite differences, for the covariance.
The result is a double approximation, the Laplace Gaussian in place of the sampled posterior and a linear map in place of \(\rvemap\), and it holds only where the forward map is smooth and near-affine.
The hierarchical posterior admits no comparable single-mode Gaussian --- \cref{sec:laplace} shows it has no mode in the first place --- and we propagate it by \MC\ alone, as described next.

\subsection{Propagating the hierarchical predictive}
\label{sec:propagate-hierarchical}

The hierarchical predictive of \cref{eq:hier-predictive} admits aggregate-to-aggregate variation, and we let the \RVE\ inherit that variation by assigning each aggregate its own descriptor.
Each \MC\ iteration realizes one \RVE, and within it the \ITZ\ around every aggregate is a draw from the same material, so we draw the population pair \(\poppar\) once per iteration and only the descriptors independently within it,
\begin{equation}
  \poppar \sim \probabilityof[\alldata]{\popmean, \popcov},
  \qquad
  \newdesc_f \mid \poppar \sim \normal{\popmean}{\popcov}, \quad f = 1, \dots, \naggr,
  \label{eq:hier-rve-draw}
\end{equation}
and map each \(\newdesc_f\) to the physical descriptor \(\params_f = (\modratio_f, \lengthscale_f, \modulus_{\infty,f})\) through the transform of \cref{sec:prior}.
Sharing one \(\poppar\) draw across the \(\naggr\) aggregates of an \RVE, rather than redrawing it per aggregate, is deliberate.
Given \(\poppar\), the descriptors \(\newdesc_{1:\naggr}\) of \cref{eq:hier-rve-draw} are independent, but \(\poppar\) is itself uncertain, and once it is marginalized out the aggregates become dependent: a draw that places the population mean \(\popmean\) low places every aggregate of the \RVE\ low together.
Drawing each aggregate independently from the predictive of \cref{eq:hier-predictive} would reproduce the distribution of every single aggregate but discard this dependence.
Because \(\qoi\) depends on the whole configuration of the \RVE\ rather than on one aggregate at a time, the discarded dependence matters: in the homogenized stress \(\macrostressaxial\), for instance, the uncertainty shared through \(\popmean\) would average out across the aggregates as though it were aggregate-to-aggregate scatter, and the uncertainty in \(\qoi\) would be understated.

Each aggregate's \ITZ\ now defines its own recovery field,
\begin{equation}
  \modulus_f(\vx) = \modulus_{\infty,f} \left[ 1 - (1 - \modratio_f)\, \exp\!\left( -\frac{\distfield_f(\vx)}{\lengthscale_f} \right) \right],
  \label{eq:per-fiber-recovery-field}
\end{equation}
with \(\distfield_f(\vx)\) the distance from \(\vx\) to the surface of aggregate \(f\), and these per-aggregate fields must be combined into a single matrix modulus.
Assigning each point to its nearest aggregate would partition the paste into Voronoi cells, but because the bulk modulus \(\modulus_{\infty,f}\) is inferred per aggregate and generally differs between neighbours, such a hard assignment makes \(\modulus(\vx)\) jump across the midplanes between aggregates.
We remove these spurious discontinuities by blending the fields with distance-based weights, taking the softmin over the aggregates,
\begin{equation}
  \modulus(\vx) = \sum_{f=1}^{\naggr} \blendweight_f(\vx)\, \modulus_f(\vx),
  \qquad
  \blendweight_f(\vx) = \frac{\exp\!\left( -\distfield_f(\vx) / \blendtemp \right)}{\sum_{h=1}^{\naggr} \exp\!\left( -\distfield_h(\vx) / \blendtemp \right)},
  \label{eq:softmin-blend}
\end{equation}
where \(\blendtemp\) sets the sharpness of the transition: the larger \(\blendtemp\), the smoother the blending.
The blended field is then used in place of \cref{eq:rve-recovery-field}, and the solve and quantity extraction proceed as in the pooled case.
\Cref{fig:rve-efield-per-agg} shows a modulus field assembled in this way; the aggregates relax to visibly different bulk moduli, in contrast to the single bulk value of the pooled field.

\section{Numerical experiments and results}
\label{sec:results-itz}

We now report the inferred \ITZ\ descriptors for the pooled and hierarchical model and report the results of propagating them through the mesoscale \RVE\ model.

\subsection{Posterior inference}
\label{sec:posterior-inference}

\begin{figure}[tbp]
  \centering
  \subcaptionbox{\label{fig:theta-marginals}}[\textwidth]{%
    \includegraphics[width=\textwidth]{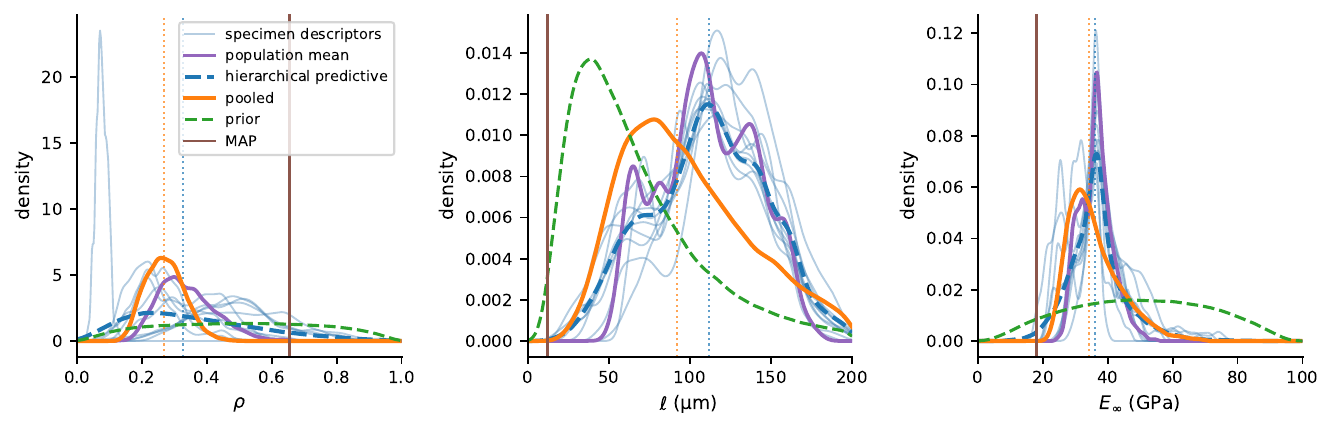}}

  \subcaptionbox{\label{fig:psi-marginals}}[\textwidth]{%
    \includegraphics[width=\textwidth]{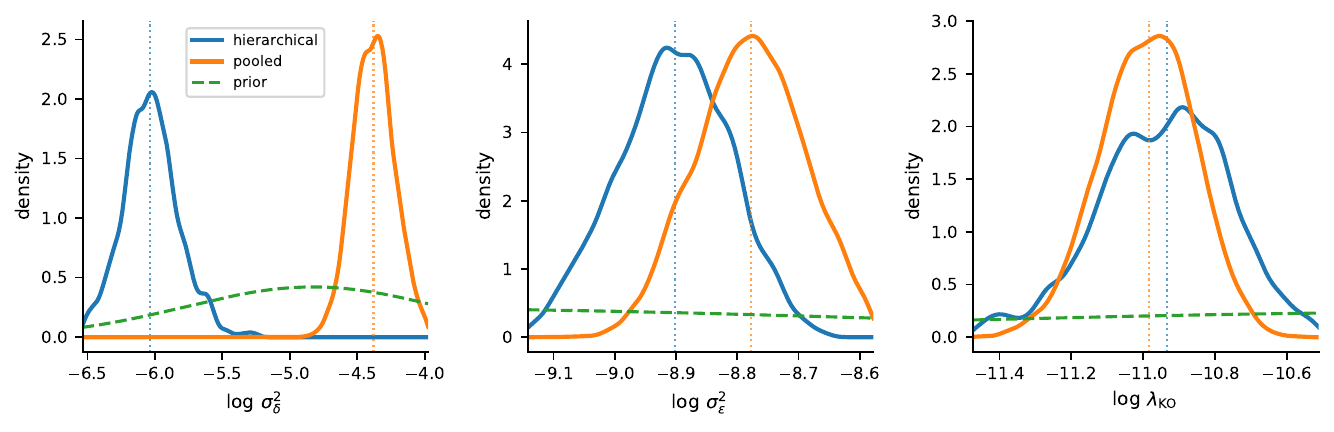}}
  \caption{
    Prior and posterior marginals of the \ITZ\ descriptors \(\params = (\modratio, \lengthscale, \Einf)\) in a and the hyperparameters \(\hyperparams\) in b --- for the pooled posterior of \cref{eq:posterior} in orange, the hierarchical predictive \cref{eq:hier-predictive} in blue, and the MAP point in brown.
    The dotted vertical lines mark the respective means of the distributions.
  }
  \label{fig:marginals}
\end{figure}

\Cref{fig:marginals} shows the prior and posterior marginals of the \ITZ\ descriptors \(\params = (\modratio, \lengthscale, \Einf)\) and the hyperparameters \(\hyperparams\).
The figure features the posterior from the pooled model of \cref{eq:posterior} alongside the prior and the \MAP\ point.
For the hierarchical model of \cref{eq:hier-model} we display the per specimen descriptors, the population mean, and the hierarchical predictive.

Comparing the pooled posterior to the prior distribution for the ITZ descriptors, we observe that the pooled posterior contracts strongly around \(\modratio\) and \(\Einf\), whereas the marginal of the length scale \(\lengthscale\) remains comparatively broad.
The hyperparameters likewise contract strongly, and the posterior mean of the noise variance \(\noisevar\) lies close to its true value.
The sharp peaks of the discrepancy amplitude \(\sigma_{\delta}\) and the correlation length \(\corrlength\) indicate that the data are informative about the scale of the model error.

Turning to the hierarchical model, the specimen marginals contract more strongly than the pooled posterior.
At the same time they vary considerably from specimen to specimen, reflecting the differences in microstructure.
The population mean itself is similar in shape to the pooled posterior.
The hierarchical predictive reproduces almost the same means as the pooled posterior for the stiffness ratio, at \(\modratio \approx 0.3\), and the far-field stiffness, at \(\Einf \approx \SI{37}{GPa}\).
It differs more in the length scale, where the hierarchical predictive mean of \(\lengthscale \approx \SI{115}{\micro\meter}\) exceeds the pooled value of \(\lengthscale \approx \SI{90}{\micro\meter}\).
Where the hierarchical predictive differs is the variance of the descriptors: it is wider than the pooled posterior and the population mean.
This increased variance is a result of the marginalization over the population spread, reflecting the between-specimen variability.

Looking at the inferred hyperparameters, the correlation length \(\corrlength\) shows almost identical sharp peaks for both models.
The noise variance \(\noisevar\) shows strong posterior contraction as well, with the hierarchical model's posterior mean is slightly lower than the pooled model's, but both remain close to the true value.
The discrepancy amplitude \(\discvar\) is markedly larger for the pooled model at \( \log \discvar \approx \num{-4.5}\) than for the hierarchical model at \( \log \sigma_{\delta}^2 \approx \num{-6.0}\).
This is to be expected: in the pooled posterior the discrepancy term absorbs the differences between specimens.
In the hierarchical model, in contrast, the discrepancy accounts only for the model error of each individual specimen, and the between specimen variability is captured by the population spread \(\popcov\).

The \MAP\ point is also indicated in the plots, and it sits far from the pooled posterior mean, at a larger \(\modratio\) and a smaller \(\Einf\) and \(\lengthscale\).
This separation is a clear indication that model discrepancy is present and that the \MAP\ point is a flawed parameter estimate.

\begin{figure}[tbp]
  \centering
  \includegraphics[width=\textwidth]{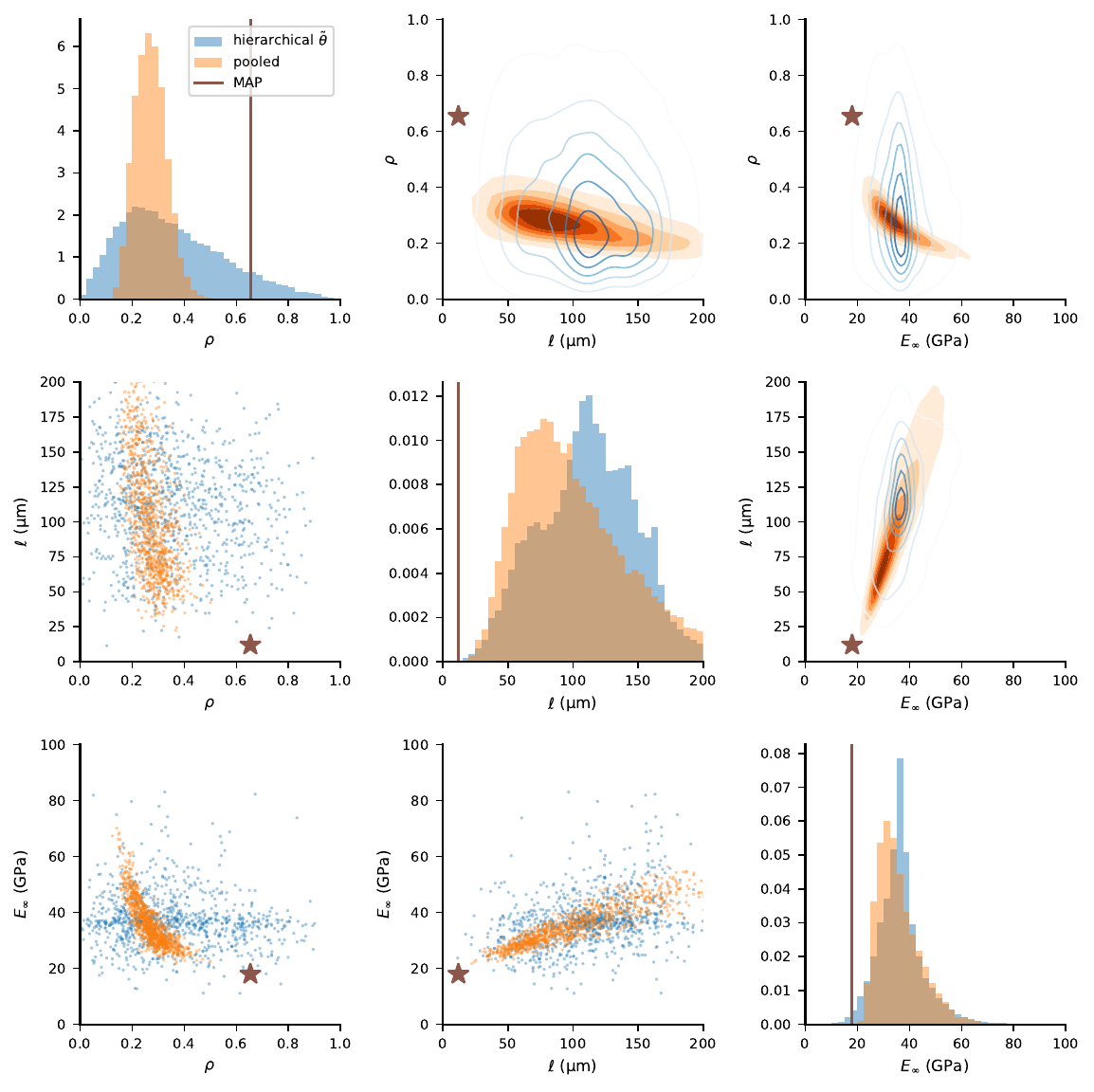}
  \caption{Pair-wise joint posterior distributions of the \ITZ\ descriptors \(\params = (\modratio, \lengthscale, \Einf)\)}
  \label{fig:posterior-pairplot}
\end{figure}

\Cref{fig:posterior-pairplot} shows pairwise joint distributions, revealing the dependence between the descriptors.
The joint distributions confirm that the pooled posterior is unimodal and exhibits strong correlations between the descriptors.
The hierarchical predictive is unimodal as well, but shows weaker correlation and a wider spread than the pooled posterior.
The pooled posterior sits roughly in the middle of the hierarchical predictive, indicating that the geometric variability between specimen is not fully reflected in the parameter estimates.

The \MAP\ point, finally, lies in a low-probability region of the pooled posterior and is not representative of either distribution.
It sits at the tail of the band that the pooled posterior occupies.
Without the discrepancy term, the likelihood treats every sensor as an independent observation with the small measurement-noise variance.
It therefore becomes so sharply peaked that the prior has little regularizing influence, and the point estimate settles wherever the misspecified model best fits the data.
The discrepancy-aware likelihood, in contrast, attributes the spatially smooth part of the misfit to \(\delta\) rather than to \(\params\).

\begin{figure}[tbp]
  \centering
  \subcaptionbox{\label{fig:prior-posterior-fields}}{%
    \includegraphics[width=.5\textwidth]{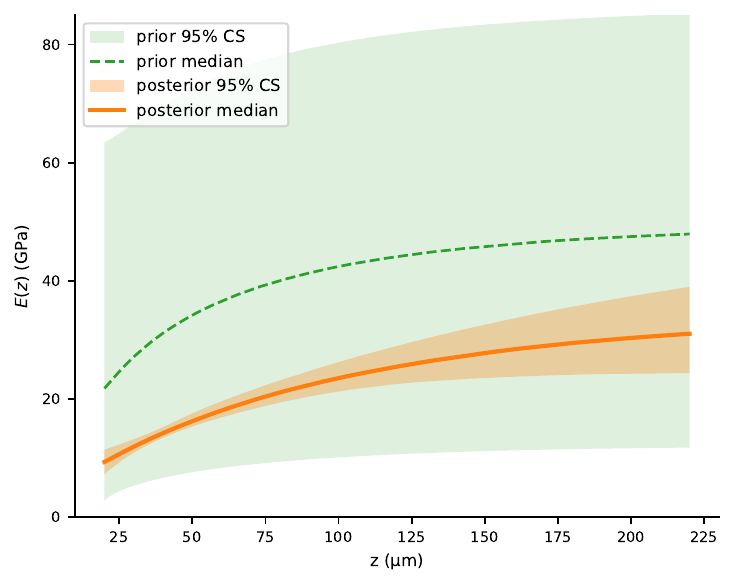}}%
  \subcaptionbox{\label{fig:posterior-field-comparison}}{%
    \includegraphics[width=.5\textwidth]{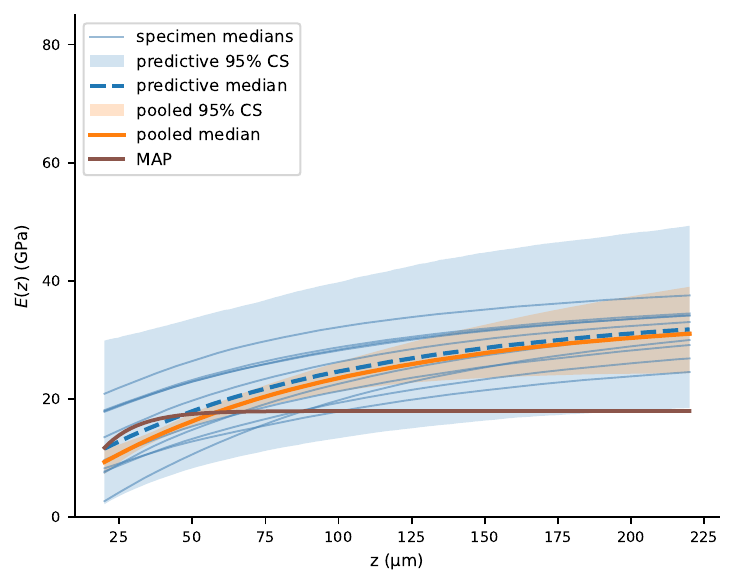}}%
  \caption{
    Prior (green) and pooled posterior (orange) in (a), and comparison of the pooled posterior and hierarchical predictive in (b), alongside the MAP point (brown).
    The pooled posterior produces a slightly softer interface on average than the hierarchical predictive.
  }
  \label{fig:posterior-fields}
\end{figure}

The posterior can also be viewed in the physical space of the modulus field \(\modulus(\axial)\), obtained by pushing the posterior samples through the forward map of \cref{eq:recovery-field}.
\Cref{fig:prior-posterior-fields} shows the prior and the pooled posterior in this space, where the strong contraction around a compliant interface is evident:
although the deliberately broad prior covers a wide range of stiffness gradients, the posterior concentrates on a much narrower range of softer interfaces, confirming that the data are informative about the \ITZ\ stiffness.

\Cref{fig:posterior-field-comparison} compares the fields resulting from the pooled posterior with those from the hierarchical predictive.
The two agree closely on the mean, but the pooled posterior yields a markedly narrower band.
The hierarchical predictive also captures the spread of the per specimen medians well, in contrast to the pooled posterior, whose band is noticeably narrower than the variability the individual specimens exhibit.
The gap between the two bands is largest for small \(\axial\), i.e.\ closest to the interface, which is precisely the region the \ITZ\ model is meant to characterize.
Getting the distribution right there matters most.

The \MAP\ point, although it lies in a low-probability region of the pooled posterior.
This is not a numerical fluke: we restarted the optimization with ten different initializations, all of which yielded the same result.
The \MAP\ point produces a modulus field in the vicinity of the pooled posterior close to the interface, but it deviates from both posteriors further away.
The shape is markedly different from both probabilistic models, with a slightly stiffer interface and a much more rapid recovery to a lower bulk modulus.
This is consistent with the model being the least sensitive to the stiffness at the free end of the beam, where that segment contributes little to the overall response and hence has little impact on the likelihood.

In conclusion, the observed differences between the results from the different approaches support the case for the hierarchical model over the pooled posterior.
By separating the between-specimen spread \(\popcov\) from the discrepancy \(\discvar\), it avoids conflating microstructural variability with model error, whereas the pooled posterior drives both into the same term.
And because the pooled posterior sits inside the hierarchical predictive rather than spanning it, treating the pooled posterior alone as the material's uncertainty would understate the geometric variability the specimens actually exhibit.
We therefore treat the hierarchical predictive, not the pooled posterior, as the more faithful characterization of the material.
We nonetheless carry the pooled posterior forward alongside it into the mesoscale propagation in the following, to show the implications of this modelling choice on the mesoscale response.

\subsection{Forward Propagation of the Posterior}\label{subsec:forward-propagation-of-the-posterior}
We propagate the posterior through the mesoscale \RVE\ model of \cref{sec:rve-model}.
We first compare the \MC\ propagation of the posterior samples with the Laplace approximation obtained by linearization, for the pooled posterior, to expose the effect of a Gaussian approximation, and then compare the pooled posterior with the hierarchical model.
Every \MC\ propagation uses all \num{20000} samples from the discretized \BPS\ trajectory.

\begin{figure}[tbp]
    \centering
    \includegraphics[width=\textwidth]{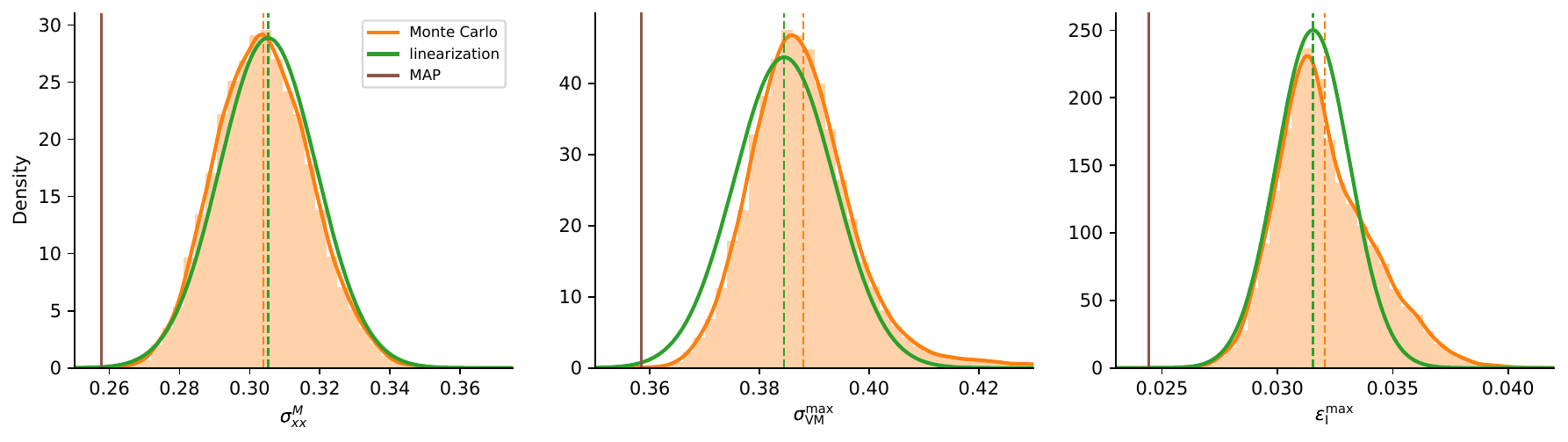}
    \caption{
        Forward propagation of the \BPS-samples via \MC\ (orange), the Laplace approximation via linearization (green), and the MAP point (brown).
        The dashed lines mark the respective empirical means.
        The linearization exhibits slightly larger variance and a shifted mean compared to the full \MC\ approach.
        The \MAP\ point is a poor predictor of all three quantities of interest.
    }
    \label{fig:forward-propagation-unscented}
\end{figure}

\Cref{fig:forward-propagation-unscented} compares the \MC\ propagation of the pooled posterior with the Laplace approximation obtained by linearization for the three quantities of interest, the macroscopic axial stress \(\macrostressaxial\), the maximum von Mises stress \(\maxvonmises\), and the maximum principal strain \(\maxprincipal\).
The linearization reproduces the mean of the maximum principal strain well, but shows errors in the mean macroscopic axial stress and the maximum von Mises stress.
It predicts the correct variance of the quantities of interest, but fails to capture the skewness of the distributions relative to the full \MC\ propagation.
The \MAP\ point is a poor predictor of all three quantities, lying far from the mean of the \MC\ propagation and outside the support of the distributions.

\begin{figure}[tbp]
  \centering
  \includegraphics[width=\textwidth]{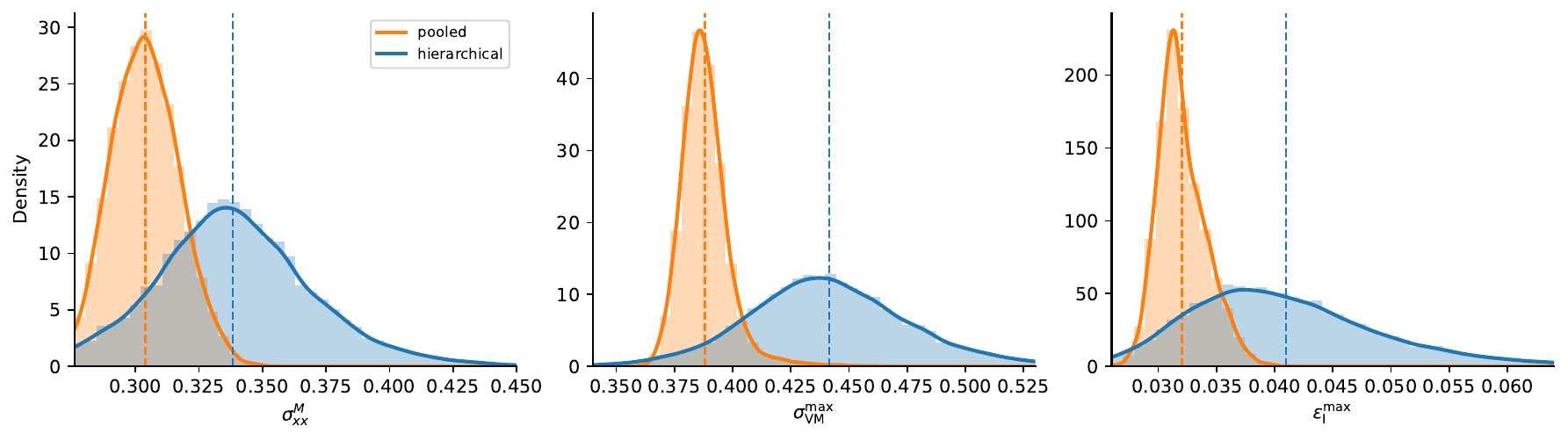}
  \caption{
    Forward propagation of the pooled (orange) and hierarchical (blue) posteriors to macroscopic stress \(\macrostressaxial\), maximum von Mises stress \(\maxvonmises\), and maximum principal strain \(\maxprincipal\).
    The dashed lines mark the respective empirical means.
    The pooled posterior produces lower means and variances than their hierarchical equivalent for all quantities.
  }
  \label{fig:forward-propagation-comparison}
\end{figure}

We now compare the push forward distributions produced by the pooled posterior with the hierarchical posterior \cref{fig:forward-propagation-comparison}.
The hierarchical posterior yields larger means for all three quantities, the macroscopic axial stress, the maximum von Mises stress \(\maxvonmises\), and the maximum principal strain \(\maxprincipal\), than its pooled counterpart.
The hierarchical posterior's variance is also larger than that of the pooled posterior, as expected given the between-specimen variability that the pooled posterior cannot capture.
The larger stresses of the hierarchical model are consistent with the posterior fields of \cref{fig:posterior-field-comparison}, where it produces a stiffer interface than the pooled model and hence a higher stress level under the prescribed macroscopic strain.
The same shift also raises the maximum principal strain: because the \RVE\ is strain-controlled, the average strain is fixed and the local strain redistributes according to the stiffness contrast rather than the absolute stiffness, so a stiffer interface sharpens the strain concentration in the compliant phase.
The gap between the two distributions is larger for the maximum von Mises stress than for the macroscopic axial stress.
The latter is a volume average and hence less sensitive to the local stiffness contrasts which are more pronounced in the hierarchical model.
For the maximum stress it matters more that different properties are assigned to the \ITZ\ around each aggregate.
The larger gap quantifies how much local fields are affected when variability in \ITZ\ properties is accounted for within the macromodel.

\begin{figure}[tbp]
  \centering
  \subcaptionbox{pooled\label{fig:max-stess-pooled}}{%
    \includegraphics[width=.4\textwidth]{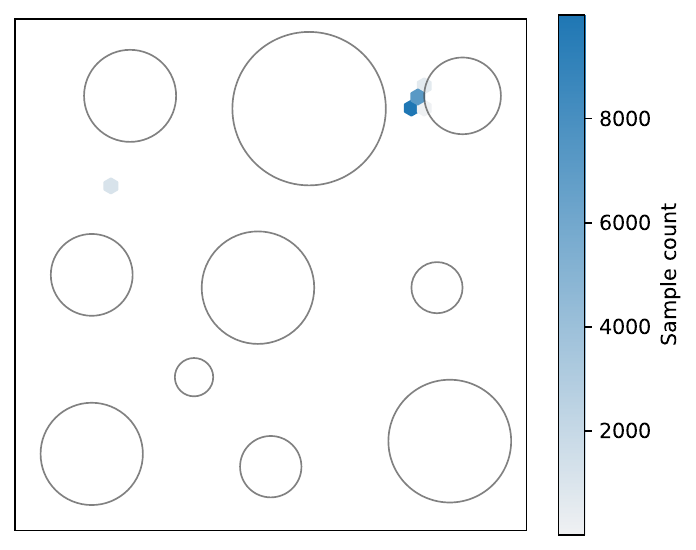}}%
  \subcaptionbox{hierarchical\label{fig:max-stress-hierarchical}}{%
    \includegraphics[width=.4\textwidth]{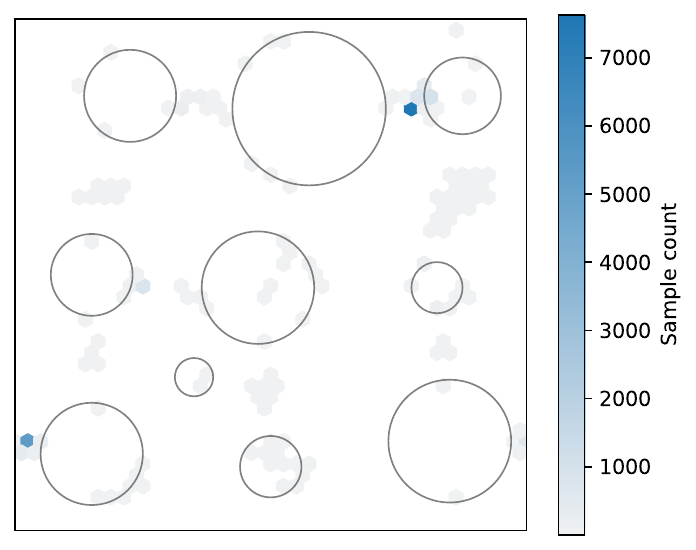}}%
  \caption{
    The locations of the maximum von Mises stress \(\maxvonmises\) for the pooled posterior (a) and the hierarchical posterior (b).
    For the pooled case, locations are concentrated in the top right, whereas the hierarchical locations are much more spread out.
  }
  \label{fig:max-stress-strain-fields}
\end{figure}

Finally, we examine the spatial localization of the maximum von Mises stress \(\maxvonmises\) in \cref{fig:max-stress-strain-fields} for the pooled and hierarchical posteriors.
The hierarchical posterior produces a wide spread of locations for the maximum von Mises stress.
In this, the larger variance in the hierarchical distribution also carries over to these locations.
The pooled posterior, in contrast, only features one single cluster on the top right of the \RVE.
While the exact Young's modulus field differs from sample to sample, the stiffness contrast throughout the domain remains similar, leading to a much more consistent mechanical response.

\FloatBarrier

\section{Conclusions and outlook}
\label{sec:conclusions}

We presented a workflow that calibrates a homogenized mesoscale model of the \ITZ\ against microscale displacement data while quantifying the uncertainty of the inferred descriptors.
The calibration is posed as a Bayesian inverse problem with a model-discrepancy term and sampled with the surrogate-assisted \BPS\ of~\cite{Riccius2026b}, marking the first application of this sampler to a real-world engineering problem.
For the hierarchical model we extended this sampler to a local \BPS\ that exploits the factorized structure of the joint posterior.
With this extension we kept the cost of a candidate event at one single finite-element solve despite the coupled population of specimens with its 39 dimensional state space.
We demonstrated the workflow on synthetic but noisy displacement data from a fully resolved finite-element model of the digital specimens of Zhang et al.~\cite{Zhang2019}.
The same workflow applies for noise-free observations, turning it into pure geometry-driven homogenization scheme.

The calibration proved robust to the observation noise: it recovered the true noise level rather than confounding it with the model discrepancy, so the inferred discrepancy reflects genuine model error.
The posterior contracts strongly from a deliberately broad prior onto a compliant interface, confirming that the data are informative about the \ITZ\ stiffness.
The modulus ratio, at \(\modratio \approx 0.3\), and the far-field stiffness, at \(\Einf \approx \SI{37}{\giga\pascal}\), are well constrained.
The recovery length scale \(\lengthscale\) remains comparatively uncertain.
Even so, its posterior places the \ITZ\ thickness above the \SIrange{30}{80}{\micro\meter} commonly assumed in the literature, a concrete correction to a widely used modelling assumption.
The \MAP\ point with a standard likelihood lies in a low-probability region of the posterior, illustrating the importance of the discrepancy term and highlighting the merits of a full probabilistic treatment over a point estimate.

The pooled and hierarchical formulations agree closely on the descriptor means but differ in what their uncertainty represents.
A single pooled descriptor vector cannot express the aggregate-to-aggregate variability between specimens, so the pooled posterior absorbs it into an inflated discrepancy term and its width reflects only estimation uncertainty.
The hierarchical model instead carries that variability explicitly in its population covariance, separating the estimation uncertainty that shrinks as specimens are added from the material variability that does not, and we regard its predictive as the more faithful characterization of the \ITZ.
This predictive is broader than the pooled posterior, and propagated through the \RVE\ it yields larger means and variances for the macroscopic axial stress, the maximum von Mises stress, and the maximum principal strain.
The full Monte Carlo propagation exposes the skewness and heavy tails that the commonly employed second-moment Laplace approximation of these quantities cannot capture.

The forward propagation in this work remains within the linear-elastic regime, yet the two formulations already behave markedly differently.
The pooled posterior produces a consistent mechanical response across its samples, confining the stress and strain concentrations to a single region of the \RVE.
The hierarchical predictive, by contrast, produces a wide variety of deformation patterns and, with them, of maximum von Mises stress and maximum principal strain locations.
We expect this difference to be amplified beyond the elastic regime: crack nucleation and propagation are highly sensitive to local stiffness contrasts and to where the stress and strain concentrate.
The single localization mode of the pooled posterior would yield a narrow range of failure patterns, whereas the dispersed concentrations of the hierarchical predictive would translate into a much wider variety of crack nucleation sites and propagation paths.

These expectations remain assumptions: the present study, confined to the elastic regime, is intended as a motivation to visit the non-linear regime, where they can be tested directly.
More broadly, it suggests that modelling aggregate and paste as a two-phase composite --- or a three-phase composite once the \ITZ\ is represented explicitly --- with constant properties within each phase is not an adequate description of the material, and that the spatial variability revealed by the hierarchical calibration must be carried into the mesoscopic prediction.
While the present study presents one way to retain the spatial variability of the \ITZ\ in a homogenized model, other approaches, that include a circumferential gradient of the stiffness around inclusions, may be even more accurate and should be explored in future work.
Modelling variation in properties along a single aggregate would require information on the spatial distance between the tested micro-cantilever and a probabilistic model that includes spatial correlation.
Applying the workflow to physical micro-cantilever measurements, rather than the synthetic data used here, is the natural next step.
Because the workflow needs only displacement data from specimens in which the \ITZ\ dominates the response, it is not tied to Portland cement and river gravel.
Recalibratig the \ITZ\ descriptors for alternative binders and for recycled aggregates would follow the same procedure and yield a more complete understanding of these increasingly adopted materials.
Finally, a richer set of specimen geometries and loading conditions, bending and torsion among them, would probe the \ITZ\ more fully and yield a comprehensive characterization of its mechanical behaviour.

\section*{Acknowledgements}
This work is supported by the TU Delft AI Labs programme through the SLIMM lab.

\section*{Code availability}
The code used to generate the results of this study is available at \href{https://github.com/SLIMM-Lab/pdmp}{\nolinkurl{github.com/SLIMM-Lab/pdmp}}.

\appendix
\crefalias{section}{appendix}%
\crefalias{subsection}{subappendix}%

\newcommand{\nbhd}[1]{N(#1)}                       %
\newcommand{\massmat}{\mM}                          %
\newcommand{\deltarate}[1][]{\delta_{\lambda#1}}    %
\newcommand{\bounded}{\hat{U}}                       %

\section{A local bouncy particle sampler for the hierarchical posterior}
\label{app:local-bps}

The hierarchical model of \cref{sec:hierarchical} leaves a \num{39}-dimensional joint posterior over the specimen descriptors \(\descl_{1:\nspec}\), the population parameters \(\popmean\) and \(\log \popsds\), and the shared log-hyperparameters \(\log \hyperparams\).
The pooled sampler of \cref{sec:sampling} does not carry over unchanged: it fits one surrogate to a single potential and simulates one Poisson clock, whereas the hierarchical target couples ten forward models through the population and, evaluated as a monolithic density, would cost \(\nspec\) finite-element solves per event.
This appendix sets out the local \BPS\ that avoids this cost, exploiting the structure that the target is not a \num{39}-dimensional black box but a sparse factor graph.
It reuses the surrogate-assisted thinning of~\cite{Riccius2026b}, applied per factor, and the local bouncing construction of~\cite{Bouchard-Ct2018}.

\subsection{The factor graph}
\label{app:factor-graph}

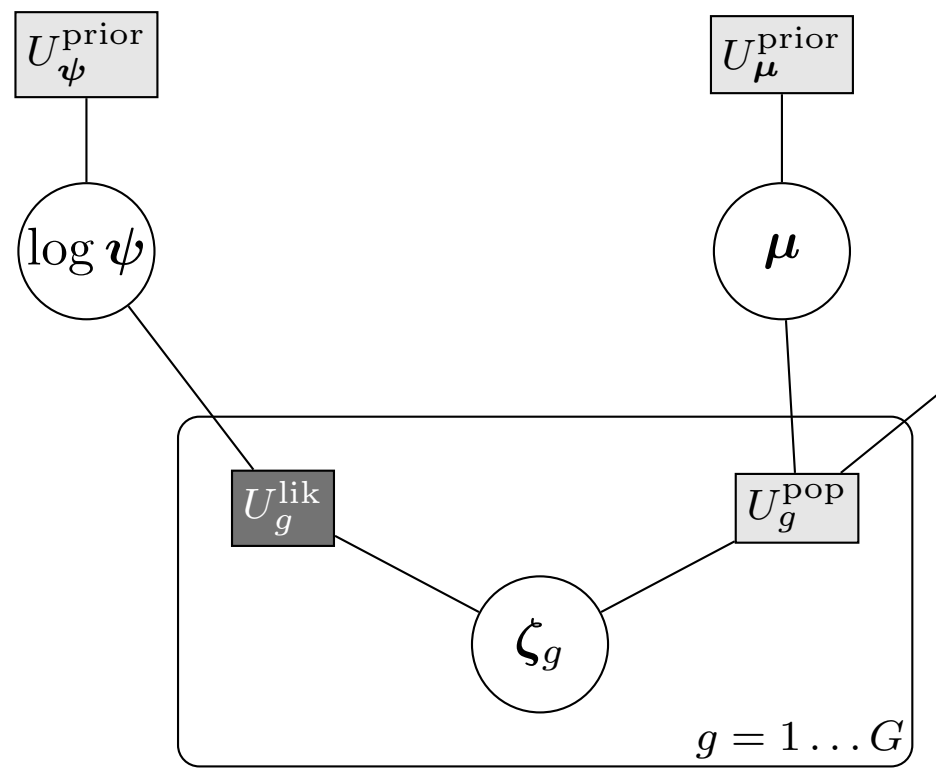
\begin{figure}[tbp]
  \centering
  \begin{tikzpicture}[
      var/.style={circle, draw, minimum size=9mm, inner sep=0pt, font=\small},
      fac/.style={rectangle, draw, inner sep=2pt, font=\scriptsize, fill=black!10},
      exp/.style={fac, fill=black!55, text=white},
      pl/.style={draw, rounded corners, inner sep=3.5mm},
      lbl/.style={font=\scriptsize, inner sep=1.5pt},
    ]
    \node[var] (psi) at (-2.7,2.6) {\(\log \hyperparams\)};
    \node[var] (mu)  at ( 1.9,2.6) {\(\popmean\)};
    \node[var] (lg)  at ( 4.1,2.6) {\(\log \popsds\)};
    \node[fac] (ppsi) at (-2.7,3.9) {\(U^{\mathrm{prior}}_{\hyperparams}\)};
    \node[fac] (pmu)  at ( 1.9,3.9) {\(U^{\mathrm{prior}}_{\popmean}\)};
    \node[fac] (plg)  at ( 4.1,3.9) {\(U^{\mathrm{prior}}_{\popsds}\)};
    \draw (ppsi) -- (psi);
    \draw (pmu)  -- (mu);
    \draw (plg)  -- (lg);
    \node[exp] (lik) at (-1.4,0.9) {\(U^{\mathrm{lik}}_g\)};
    \node[var] (xi)  at ( 0.3,0.0) {\(\descl_g\)};
    \node[fac] (pop) at ( 2.0,0.9) {\(U^{\mathrm{pop}}_g\)};
    \node[pl, fit=(lik)(xi)(pop),
          label={[lbl, anchor=south east]south east:\(g = 1 \dots \nspec\)}] (plate) {};
    \draw (lik) -- (xi);
    \draw (lik) -- (psi);
    \draw (pop) -- (xi);
    \draw (pop) -- (mu);
    \draw (pop) -- (lg);
  \end{tikzpicture}
  \caption{
    The hierarchical target of \cref{eq:lbps-potential} as a factor graph.
    Circles are variable blocks and boxes are factors; the dark box which houses \(U^{\mathrm{lik}}_g\) is the only expensive one, costing a finite-element solve per gradient.
    The likelihood factor does \emph{not} read \(\popmean\) or \(\log \popsds\): its neighbourhood is \((\descl_g, \log \hyperparams)\) alone, which keeps the graph sparse and lets each variable block be preconditioned by a disjoint set of factors.
  }
  \label{fig:lbps-factor-graph}
\end{figure}

The potential --- the negative log-posterior --- decomposes into terms that each depend on a small subset of the state,
\begin{equation}
  \begin{split}
    U(\descl_{1:\nspec}, \popmean, \log \popsds, \log \hyperparams)
    &= \underbrace{\sum_{g=1}^{\nspec} U_g^{\mathrm{lik}}(\descl_g, \log \hyperparams)}_{\text{one solve each}}
    + \sum_{g=1}^{\nspec} U_g^{\mathrm{pop}}(\descl_g, \popmean, \log \popsds) \\
    &\quad + U^{\mathrm{prior}}_{\popmean} + U^{\mathrm{prior}}_{\popsds} + U^{\mathrm{prior}}_{\hyperparams}.
  \end{split}
  \label{eq:lbps-potential}
\end{equation}
For \(\nspec = 10\) this is \num{23} factors, of which only the ten likelihood factors \(U_g^{\mathrm{lik}}\) are expensive, each costing one reverse-mode finite-element solve for its gradient.
No new forward model is needed: \(U_g^{\mathrm{lik}}\) is the per-specimen potential built from the likelihood of \cref{eq:per-specimen-likelihood} with its prior removed.
\Cref{fig:lbps-factor-graph} shows the graph and \cref{tab:lbps-factors} lists the factors with their neighbourhoods \(\nbhd{f}\), the set of coordinates factor \(f\) reads.
Sampled in \(\descl_g\) directly, each likelihood factor reads only its own three descriptors and the three shared log-hyperparameters \(\log \hyperparams\), and the population level is written as \(\nspec\) separate factors so that a change to \(\descl_g\) invalidates only its own population term.

\begin{table}[tbp]
  \centering
  \caption{The \num{23} factors of \cref{eq:lbps-potential} for \(\nspec = 10\), with their neighbourhoods and the Poisson clock each one receives (\cref{app:clocks}).}
  \label{tab:lbps-factors}
  \begin{tabular}{@{}llccl@{}}
    \toprule
    Factor & Neighbourhood \(\nbhd{f}\) & Count & Gradient cost & Clock \\
    \midrule
    \(U^{\mathrm{lik}}_g\) & \((\descl_g, \log \hyperparams)\) & \num{10} & one solve & analytic \\
    \(U^{\mathrm{pop}}_g\) & \((\descl_g, \popmean, \log \popsds)\) & \num{10} & closed form & integrated \\
    \(U^{\mathrm{prior}}_{\popmean}\) & \(\popmean\) & \num{1} & closed form & analytic \\
    \(U^{\mathrm{prior}}_{\popsds}\) & \(\log \popsds\) & \num{1} & closed form & integrated \\
    \(U^{\mathrm{prior}}_{\hyperparams}\) & \(\log \hyperparams\) & \num{1} & closed form & analytic \\
    \bottomrule
  \end{tabular}
\end{table}

\subsection{Bouncing locally}
\label{app:local-bounce}

Under the pooled posterior the homogenized model shares one geometry across specimens, so a single solve yields the displacement field and every specimen's residuals follow by reading it off at different sensors.
That no longer holds here, because each \(\descl_g\) induces a different modulus field, so evaluating the exact \emph{total} rate of a global \BPS\ costs \(\nspec\) solves per event.
A summed rate bound does not rescue this: the inequality \((\sum_g \innerproduct{\vv}{\nabla U_g})_{+} \le \sum_g M_g\) is valid, but thinning against it still requires the exact total rate at every candidate, that is, all \(\nspec\) solves.

The remedy is to bounce locally rather than globally.
Each factor of \cref{eq:lbps-potential} carries its own Poisson clock, running at the rate its own neighbourhood sees,
\begin{equation}
  \lambda_f(\vx, \vv) = \left( \innerproduct{\vv_{\nbhd{f}}}{\nabla U_f(\vx_{\nbhd{f}})} \right)_{+},
  \label{eq:lbps-rate}
\end{equation}
the earliest clock wins, and when factor \(f\) fires the velocity is reflected using \(\nabla U_f\) alone, leaving every coordinate outside \(\nbhd{f}\) untouched,
\begin{equation}
  \vv_{\nbhd{f}} \gets \vv_{\nbhd{f}} - 2 \frac{\innerproduct{\vv_{\nbhd{f}}}{\nabla U_f}}{\norm{\nabla U_f}^2} \nabla U_f.
  \label{eq:lbps-bounce}
\end{equation}
This is the local \BPS\ of~\cite{Bouchard-Ct2018}, and it is exact for the factorized target rather than an approximation of the global sampler: \cref{eq:lbps-bounce} is self-inverse and preserves \(\norm{\vv}\), and \(\lambda_f\) minus its value after reflection is precisely the contribution of factor \(f\) to the flow of the potential, so that summing over factors recovers the global bounce condition.
Its consequence here is that a candidate event only ever requires the factor that generated it, so the per-event cost returns to one forward solve regardless of \(\nspec\).
Velocity refreshment stays global, at the constant rate \(\refreshrate\) of \cref{sec:sampling}.

\subsection{Poisson clocks}
\label{app:clocks}

Two bookkeeping facts decide whether the local construction pays off.
The first is which clocks survive an event.
Because the flow is linear between events, a factor's clock stays valid for as long as the velocity it reads is unchanged, so only the factor that fired and its graph neighbours need re-simulating.
This is where the sparsity of \cref{fig:lbps-factor-graph} pays off: over the run an event re-drew a mean of \num{5.08} of the \num{23} clocks, where a global sampler would re-draw all of them.
The saving is bounded by the shared blocks: every likelihood factor neighbours every other one through \(\log \hyperparams\), so a bounce on a likelihood factor is not local, and it is the population and prior factors, which do not touch \(\log \hyperparams\), that are spared.

The second is that the two families of factors are drawn by different mechanisms, and this is the one that carries the cost.
The population and prior factors have closed-form gradients, so their clocks are simulated directly against the exact rate --- some by solving a quadratic equation for the first arrival when the rate is affine along the flow, the rest by numerical integration, but always exactly.
The likelihood factors instead run against a cheap surrogate bound: its first arrival gives a candidate event, and only then is the true rate, which entails one finite-element solve, evaluated to decide whether to accept it, the thinning construction of \cref{app:thinning}.
It is this thinning check, not the clock draw itself, that spends the budget: over the run, \num{78371} of these candidate checks touched a finite-element solve (\cref{tab:lbps-run}).

\subsection{Per-specimen surrogates}
\label{app:surrogates}

Each likelihood factor is simulated by thinning against a surrogate, and the exact factor is consulted only at candidate events, one true gradient per candidate against roughly a hundred for integrating the true rate.
On a forward model those are finite-element solves, and that ratio is the point of the surrogate.
The surrogate is a quadratic model of the specimen's \emph{log-likelihood} at a reference point, with input \((\descl_g, \log \hyperparams) \in \R^6\), so the fit dimension is \(3 + 3\) independently of \(\nspec\) and no surrogate over the full \num{39}-dimensional state is ever required.
Fitting the log-likelihood rather than the whole factor keeps the population term analytic and swappable, and the parameterization's chain rule is applied to the surrogate by the same code that applies it to the truth.

Two details of the quadratic fit are not incidental:
The expansion point comes from a \emph{regularized} separate fit, because the bare specimen likelihood has no usable mode: \(\discvar\) and \(\noisevar\) are identified only through their sum, and \(\Einf\) enters through a saturating transform, so an unregularized fit runs to the boundary.
The global \MAP\ is not an option either, for the same reason \cref{sec:point-estimate} withholds it from the hierarchical model: the joint density has no maximizer.
Each specimen is therefore expanded about the mode of its own posterior instead, and this moment-matched separate fit, which also serves as the sampler's starting state, stands in for the joint \MAP\ that does not exist.

Further, the fitted curvature is floored.
The measured Hessians are indefinite --- across the ten specimens \num{18} of \num{60} curvature directions came out below the floor, the worst at \num{-18.9} --- and an indefinite quadratic is a saddle, unbounded above and so a structurally wrong shape for a thinning bound, its rate falling exactly where that of the truth eventually rises.
Flooring the eigenvalues leaves value and gradient at the expansion point untouched, so the model stays tangent, and changes only the curvature, which the thinning corrects for in any case.
The cost of all this is negligible: ten surrogates, in six dimensions, fitted from \num{140} true likelihood evaluations in total.

\subsection{Thinning and adaptive offsets}
\label{app:thinning}

The correction of~\cite{Riccius2026b} carries over per factor rather than globally.
Factor \(f\) is simulated against the bounding rate \((\innerproduct{\vv}{\nabla \bounded_f})_{+} + \deltarate[,f]\), each likelihood factor with its own surrogate \(\bounded_f\) and its own offset \(\deltarate[,f]\), and each offset is raised only when its own factor's bound is caught below its own true rate.
Offsets also decay on their own factor's clock rather than globally: a global decay would change every bound at every event and so invalidate every clock, destroying the locality the sampler exists to exploit.

Thinning is exact only while the bound dominates the true rate over the whole simulated interval, not merely at the candidate point, and a surrogate offers no such guarantee.
When a violation occurs the run takes the bounce, accepting with probability \(\min(1, m/M)\) on the grounds that a true rate \(m\) exceeding the bound \(M\) wanted the event at least as much as the bound proposed it.
A small offset decay \(\beta = 0.01\) was used to allow the offset to recover after encountering an outlier.

\subsection{Mass preconditioning}
\label{app:mass}

Besides the surrogate, the geometry of the refreshment has a strong effect on the sampler's performance.
The \BPS\ refreshes its velocity isotropically, so on an ill-conditioned target it keeps re-probing whichever direction has the largest curvature, and those excursions drive both the event rate and the thinning offsets.
A diagonal mass matrix \(\massmat\) removes this: the velocity is refreshed as \(\vv = \massmat^{1/2} \vw\) with \(\vw\) uniform on the unit sphere, and reflections are taken against \(\massmat\) rather than the Euclidean metric,
\begin{equation}
  \vv_{\nbhd{f}} \gets \vv_{\nbhd{f}} - 2 \frac{\innerproduct{\vv_{\nbhd{f}}}{\nabla U_f}}{\innerproduct{\nabla U_f}{\massmat_{\nbhd{f}} \nabla U_f}} \massmat_{\nbhd{f}} \nabla U_f.
  \label{eq:lbps-mass-bounce}
\end{equation}
This is exactly the local \BPS\ run on \(\massmat^{-1/2} \vx\) with isotropic refreshment and Euclidean reflection, pushed back through that linear, constant-Jacobian map; it is a change of variables and not an approximation, which is why it leaves the affine-rate detection of \cref{app:clocks} untouched --- that test only ever sees \(\vv\) and \(\nabla U_f\), never \(\massmat\).
This achieves the same effect as the affine transformation of~\cite{Riccius2026b}, but from the other side: there the state is reparameterized by a full Cholesky factor and the sampler dynamics are left isotropic, whereas here the state is left alone and the refreshment and bounce are reparameterized instead, by a \(\massmat\) restricted to diagonal so that each factor still reads only its own neighbourhood \(\nbhd{f}\) --- a dense \(\massmat\), like a full Cholesky factor, would mix coordinates across factors and destroy the sparsity the local \BPS\ depends on.

The entries are set from an estimate of the diagonal precision at the expansion point, \(m_i = 1/P_{ii}\), where \(P_{ii}\) sums the curvature of every factor that reads coordinate \(i\), since that sum is what a bounce on any one of them sees.
The three blocks draw on disjoint sets of factors, which is what lets the diagonal mass below be estimated cleanly per block.
For \(\descl_g\), only specimen \(g\)'s own likelihood and its own population factor read it; only the likelihood term is used, because the population contribution \(1/\popsd_j^2\) swings over three orders of magnitude as \(\log \popsds\) is sampled, and freezing one arbitrary value of it into a fixed mass measures worse than omitting it.
For \(\log \hyperparams\), every specimen's likelihood factor reads it, so its mass comes from the sum of their curvatures plus the precision of its own prior.
For \(\popmean\), the likelihood factors never read it, so its curvature comes entirely from the \(\nspec\) population factors, each contributing \(1/\popsd_j^2\), plus its own prior.
On this problem the resulting mass spanned \(\left[ \num{5.0e-4}, \num{1.0} \right]\), a condition number of about \num{2.0e3}.

\subsection{Cost of a run}
\label{app:lbps-cost}

\Cref{tab:lbps-run} collects the measured anatomy of a run at horizon \(T = \num{10000}\) on the ten-specimen problem, which took about \SI{40}{\hour} of wall time, and \cref{alg:lbps} states the procedure.
The line to read is the second block: \num{78371} candidate events entailed a finite-element solve and \num{10484} of those became accepted bounces, so the hierarchical target is sampled at roughly \num{7.8} solves per unit of process time despite carrying ten forward models.

\begin{table}[tbp]
  \centering
  \caption{Measured anatomy of the local \BPS\ run at \(T = \num{10000}\), \(\nspec = 10\).}
  \label{tab:lbps-run}
  \begin{tabular}{@{}lr@{}}
    \toprule
    Quantity & Value \\
    \midrule
    State dimension & \num{39} \\
    Factors (likelihood / population / prior) & \num{10} / \num{10} / \num{3} \\
    Factors with an analytic clock & \num{12} \\
    \midrule
    Events & \num{103306} \\
    Bounces / refreshments & \num{34384} / \num{1035} \\
    \quad of which finite-element solves (proposed / accepted) & \num{78371} / \num{10484} \\
    Clocks re-drawn per event (mean, of \num{23}) & \num{5.08} \\
    \midrule
    Bound violations & \num{1254} \\
    Final offsets, likelihood factors & \numrange{0.033}{0.967} \\
    Surrogate fitting cost (total) & \num{140} evaluations \\
    \bottomrule
  \end{tabular}
\end{table}

\begin{algorithm}[tbp]
  \caption{Hierarchical inference by a local bouncy particle sampler}
  \label{alg:lbps}
  \begin{algorithmic}[1]
    \State Fit each specimen's regularized separate posterior; collect the modes into an expansion point \(\vs^{\star}\) and a starting state.
    \State Fit \(\nspec\) quadratic surrogates of the per-specimen log-likelihood at \(\vs^{\star}\), in \num{6} dimensions, flooring the curvature eigenvalues.
    \State Build the diagonal mass \(\massmat\) from the factor curvatures at \(\vs^{\star}\) (\cref{app:mass}).
    \State Assemble the state \(\vs = (\descl_{1:\nspec}, \popmean, \log \popsds, \log \hyperparams)\) of dimension \(3\nspec + 9\).
    \State Assemble the \num{23} factors of \cref{eq:lbps-potential} with their neighbourhoods; attach a surrogate bound and an offset \(\deltarate[,f]\) to each likelihood factor and leave the analytic factors exact.
    \State Draw every factor clock
    \While{\(t < T\)}
      \State Take the earliest of the factor clocks and the global refreshment clock.
      \If{refreshment}
        \State Draw \(\vv = \massmat^{1/2} \vw\), \(\vw\) uniform on the unit sphere; re-draw all clocks.
      \ElsIf{factor \(f\) is likelihood}
        \State Evaluate the true rate \(M\) and the proposed rate \(m\) of the factor
        \State Accept the bounce with probability \(\min(1, m/M)\), reflecting by \cref{eq:lbps-mass-bounce}
        \State Grow \(\deltarate[,f]\) if \(m > M\), re-draw the clocks of that factor.
      \EndIf
    \EndWhile
    \State Discretize the trajectory at evenly spaced times to obtain the draws \(\{\vs^{(i)}\}\).
    \State Report \(\probabilityof[\alldata]{\popmean}\) and \(\probabilityof[\alldata]{\popcov}\); form the predictive of \cref{eq:hier-predictive} by drawing \(\newdesc^{(i)} \sim \normal{\popmean^{(i)}}{\popcov^{(i)}}\) for each draw and mapping back to physical space.
  \end{algorithmic}
\end{algorithm}

Two further points bound how far this construction should be read.
It exploits sparsity rather than additivity, and the two are independent: exact subsampling schemes for \PDMPs, which replace \(\nabla U\) by an unbiased estimate over a random subset of the data, could in principle be layered on top of the local bounce, since one construction decides which factor a bounce reads and the other how that factor's gradient is estimated.
On this problem, however, there is little left for them to recover.
Across specimens the saving they would be reaching for is the one the local bounce already makes: a candidate event evaluates a single likelihood factor regardless of \(\nspec\), and a subsampled estimate of the total rate would reach the same cost by a noisier route.
Within a factor the unit of cost is the finite-element solve and not the datum, since one reverse-mode solve returns the gradient however many sensors are read from the resulting field, so dropping observations buys no solves.
The bound would also have to dominate over every possible subsample, which is a difficult endeavour for a \PDE\ forward map.
Further, the adaptive offsets of \cref{app:thinning} would lose their target, as a violation could no longer be told apart from subsampling noise.
The variance-reduction idea behind these schemes is nonetheless already present: expanding the potential about a reference point once and reusing that quadratic model across candidate events is itself a control variate, which is what \cref{app:surrogates} constructs.

\FloatBarrier

\end{document}